\documentclass[useAMS,usenatbib,a4paper,fleqn]{mnras}
\usepackage[T1]{fontenc}
\usepackage{ae,aecompl}

\usepackage{graphicx,lscape,amssymb}
\usepackage{natbib}
\usepackage{amsmath}
\usepackage{hyperref}
\usepackage[usenames,dvipsnames]{xcolor}
\usepackage{soul}
\usepackage{longtable}
\usepackage{lscape}
\usepackage{multirow}
\usepackage{caption}
\usepackage[usenames,dvipsnames]{xcolor}

\title{Fundamental parameter accuracy of DA and DB white dwarfs in \textit{Gaia} Data Release 2}

\author[Tremblay et al.]{P.-E. Tremblay,$^1$ E. Cukanovaite,$^1$ N. P. Gentile Fusillo,$^1$ T. Cunningham,$^1$ and  
\newauthor{M. Hollands$^1$}\\
$^1$ Department of Physics, University of Warwick, Coventry, CV4 7AL, UK}

\begin{document}
\maketitle

\label{firstpage}
\begin{abstract}
We report on a comparison of spectroscopic analyses for hydrogen (DA) and helium atmosphere (DB) white dwarfs with {\it Gaia} Data Release 2 (DR2) parallaxes and photometry. We assume a reddening law and a mass-radius relation to connect the effective temperatures ($T_{\rm eff}$) and surface gravities ($\log g$) to masses and radii. This allows the comparison of two largely independent sets of fundamental parameters for 7039 DA and 521 DB stars with high-quality observations. This subset of the {\it Gaia} white dwarf sample is large enough to detect systematic trends in the derived parameters. We find that spectroscopic and photometric parameters generally agree within uncertainties when the expectation of a single star is verified. {\it Gaia} allows the identification of a small systematic offset in the temperature scale between the two techniques, as well as confirming a small residual high-mass bump in the DA mass distribution around 11\,000-13\,000~K. This assessment of the accuracy of white dwarf fundamental parameters derived from {\it Gaia} is a first step in understanding systematic effects in related astrophysical applications such as the derivation of the local stellar formation history, initial-to-final mass relation, and statistics of evolved planetary systems.
\end{abstract}
\section{Introduction}

The atmospheric parameters of white dwarfs, the effective temperature ($T_{\rm eff}$) and surface gravity ($\log g$), play a fundamental role in the study of post-main-sequence stellar evolution \citep[see, e.g.,][]{kalirai14,rolland18}, evolved planetary systems \citep{koester14}, Galactic formation history \citep{tremblay14}, and the calibration of instruments \citep{bohlin14}. For decades, the most precise method to determine $T_{\rm eff}$ and $\log g$ has been the spectroscopic fitting of the Balmer lines in hydrogen-atmosphere DA white dwarfs \citep{bergeron92,finley97}, and the He I lines in helium-dominated DB stars \citep{beauchampetal99-1,voss07,bergeron11}. In contrast, trigonometric parallax measurements of stellar remnants or their companions can also be used to characterise $T_{\rm eff}$ and stellar radii from photometric analyses \citep{koester79,bergeron01}. The surface gravity can then be inferred by using the white dwarf mass-radius relation \citep[see, e.g.,][]{fontaine01}, with a small dependence on the assumed internal stratification \citep{romero12}. The main advantage of the photometric technique is that it can be used to fit cool DC white dwarfs with no optical transitions as well as metal- or carbon-rich remnants where spectroscopic $\log g$ determinations are difficult \citep{dufour05,dufour07}. One shortcoming is that for $T_{\rm eff} \gtrapprox 12\,000$~K, optical colours of all stellar remnants get rather insensitive to the temperature, requiring very precise photometry for that determination \citep{carrasco14}. Until now, the main limitation of the photometric analyses has been that precise parallax measurements were available for only hundreds of white dwarfs \citep{bedard17}, compared to the $\approx$ 30\,000 objects \citep{kleinman13,kepler15, kepler16,gentilefusillo18} with spectroscopy from the Sloan Digital Sky Survey \citep[SDSS;][]{sloanDR7}. 

The spectroscopic and photometric techniques have been combined to derive masses and radii independent of a theoretical mass-radius relation \citep{vauclair97,provencal98,holberg12,tremblay17,bedard17}. This method aims at understanding the internal stratification of white dwarfs, where additional constraints from asteroseismology \citep{romero12} and statistical analyses \citep{tremblay08} can also be considered. The cornerstone European Space Agency \textit{Gaia} space mission is now challenging our knowledge of white dwarfs by providing a homogeneous and extremely precise sample of photometry and parallaxes. \citet{tremblay17} used 52 directly or indirectly (wide companions) observed white dwarfs in \textit{Gaia} Data Release 1 and found that uncertainties on the spectroscopic parameters are now the dominant uncertainties on individual mass and radius determinations for bright and close white dwarfs. The authors conclude that the comparison of the photometric and spectroscopic techniques is equally a verification of the mass-radius relation and the model atmospheres (see also \citealt{genest14}). In the latter case, a number of improvements have been made in recent years for the bulk of the white dwarfs that have H- or He-dominated atmospheres, such as Stark broadening for hydrogen lines including non-ideal gas effects \mbox{\citep{tremblay09}}, neutral broadening for Lyman $\alpha$ \citep{kowalski06}, improved opacities from collision-induced-absorption \citep[CIA;][]{blouin17}, and the account of 3D convective effects for both DA and DB stars \citep{tremblay13,cukanovaite18}. It is highly desirable to compare these theoretical developments with \textit{Gaia}.

There are approximately 250\,000 white dwarfs \citep{gentilefusillo18} in the second \textit{Gaia} Data Release \citep[DR2;][]{gaiaDR2-ArXiV-1}. The sample is estimated to be $95.8{+1.3\atop-1.7}$\,percent volume-complete within 20\,pc \citep{hollands18}, and this is likely to hold up to a distance of $\approx$ 60\,pc, where the coolest disk white dwarfs become too faint to be observed \citep{gentilefusillo18}. In addition to parallaxes that have a median precision of 1.5\% within 100\,pc, \textit{Gaia} DR2 provides colours in the broad $G$, $G_{\rm BP}$, and $G_{\rm RP}$ pass bands, which can be used to constrain white dwarf parameters using the photometric technique \citep{el-badry18,alberto18,gentilefusillo18,kilic18b}. For the 20\,pc sample, \citet{hollands18} have demonstrated that the atmospheric parameters derived from \textit{Gaia} are in very good agreement with earlier photometric analyses. Furthermore, \citet{gentilefusillo18} have shown that for known single DA white dwarfs for which the monochromatic spectral energy distribution is well modeled \citep{bohlin14}, \textit{Gaia}-derived $T_{\rm eff}$ values are in very good agreement with independent photometric temperatures obtained using instead the Pan-STARRS \citep{PanSTARRS} or SDSS data sets. This suggests that \textit{Gaia} photometry is well calibrated at the precision level of other existing photometric surveys.

In this work, we present an overview of how \textit{Gaia} DR2 compares with published or existing spectroscopic parameters for single, non-magnetic DA and DB/DBA white dwarfs \citep{gianninas11,koester15,tremblay16,rolland18}. These correspond to the spectral types where both the photometric and spectroscopic techniques are deemed reliable and independent. {\it Gaia} provides such a dramatic update in sample size compared to existing parallax samples that we can now hope to identify trends as a function of $T_{\rm eff}$, $\log g$ or spectral subtypes. Our initial approach is to combine the spectroscopic atmospheric parameters, a theoretical mass-radius relation, and the \textit{Gaia} $G$ apparent magnitude to predict a parallax that can be compared with the observed value. In a further step, we also derive independent $T_{\rm eff}$ and $\log g$ values from the photometric technique using the full \textit{Gaia} data set. In all cases we present the current picture and make no attempt to improve any of the existing models, data calibration, or fitting techniques.

In Section~\ref{sec2} we present our selected spectroscopic samples and in Section~\ref{sec3} we compare existing spectroscopic distances for DA and DB/DBA stars with \textit{Gaia} parallaxes. In Section~\ref{sec4} we compare \textit{Gaia} DR2 photometric atmospheric parameters to spectroscopic analyses, followed by a discussion in Section~\ref{sec5} and a summary in Section~\ref{sec6}.

\section{White Dwarf Samples}
\label{sec2}

The all-sky and bright sample of DA white dwarfs from \citet{gianninas11} drawn from the White Dwarf Catalog of \citet{mccook99} is a benchmark to study their fundamental parameters because of the high signal-to-noise ratio (S/N) of the spectroscopic observations. The relative proximity of the objects also allows for the most precise \textit{Gaia} data set that we employ in this study. We rely on the 1D atmospheric parameters as published by the authors. We also use a sample of DA white dwarfs drawn from the \textit{Gaia}-SDSS catalogue of \citet{gentilefusillo18}. We have secured all spectra from the SDSS SkyServer\footnote{\url{https://skyserver.sdss.org/dr14/}} with the new data reduction from DR14 \citep{Abolfathi2018}, and fitted the observations using the same technique and 1D model atmospheres as those employed for our analyses of earlier data releases \citep{tremblay11,tremblay16} as well as \citet{gianninas11}. In brief, the model atmospheres rely on the Stark broadening profiles of \citet{tremblay09}, the \citet{hm88} equation-of-state, and fully account for NLTE effects \citep{tremblay11}. For all objects where we observed the so-called Balmer-line problem \citep{werner96,gianninas10} with deeper than predicted cores at H$\alpha$ and H$\beta$, we employed the NLTE models with carbon, nitrogen, and oxygen computed by \citet{gianninas10} to provide a better fit to the Balmer lines and improved atmospheric parameters \citep[see Section 4.3 of][]{tremblay11}. 

\begin{table*}
\centering
\caption{White dwarf samples
\label{WDsample} }
\begin{tabular}{llll}
\hline
\hline
Sample & Initial & Clean (non-magnetic,  & \textit{Gaia} DR2 (with \\
     &         & single WDs, S/N $>$ 20) & quality cuts, Eqs.\,1-3)  \\
\hline
DA: \citet{gianninas11} & 1265 & 1201 & 1145 \\
DA: SDSS DR1-DR7 & 12491 & 2941 & 2726$^{a}$ \\
DA: SDSS BOSS DR9-DR14 & 8650 & 3584 & 3168$^{b}$ \\
\hline
DB: \citet{rolland18} & 119 & 119 & 116 \\
DB: \citet{koester15} & 1107 & 439 & 405 \\
\hline
\multicolumn{4}{l}{$^{a}$ Spectroscopic parameters available in Table~\ref{A1}.}\\
\multicolumn{4}{l}{$^{b}$ Spectroscopic parameters available in Table~\ref{A2}.}\\
\multicolumn{4}{l}{Notes: Spectroscopic parameters are otherwise taken from the sample papers and {\it Gaia} data}\\
\multicolumn{4}{l}{from \citet{gentilefusillo18}.}
\end{tabular}
\end{table*}

From a visual inspection of the SDSS spectra and fits to the $ugriz$ photometry, we have cleaned the sample of magnetic white dwarfs, DAO stars, cool He-rich atmospheres with weak hydrogen lines, and identifiable binaries of the types DA+DC, DA+DB and DA+dM. For the \citet{gianninas11} sample, we have used the existing flags in their catalogue to filter these subtypes. We note that double DA white dwarfs can not be identified from the spectrum and photometry alone except for rare double-lined binaries \citep{tremblay11}. We kept all DAZ white dwarfs as the presence of metals does not significantly influence the atmospheric parameters. For SDSS spectra, we restrict our analysis to objects with S/N $>$ 20 since this high-quality subsample is cleaner and already large enough to provide a robust comparison with \textit{Gaia}. 

In the following, we split the SDSS spectra in two subsamples, first based on observations made with the original SDSS spectrograph \citep{sloanDR7} up to DR7. This subsample largely overlaps with the catalogue of \citet{kleinman13} apart from a small fraction of cool DA white dwarfs not covered by their colour cuts \citep{gentilefusillo18}. The second sample, overlapping with \citet{kepler15,kepler16}, is for white dwarfs observed within the newer SDSS-III Baryon Oscillation Spectroscopic Survey \mbox{\citep[BOSS;][]{sloanDR9}}, corresponding to the SDSS Data Releases 9 to 14. The adopted spectroscopic samples of DA white dwarfs are summarised in Table~\ref{WDsample}. We emphasise that by using the same models and fitting technique for all three data sets, we can more easily study the effects from the different instruments and data reduction.

We also perform a comparison of the spectroscopic parameters of DB and DBA white dwarfs with \textit{Gaia}. We rely on the all-sky sample of bright objects from \citet{rolland18} and the SDSS DR12 analysis of \citet{koester15}. In both cases we employ the published atmospheric parameters. As for the DA white dwarfs in the SDSS, we restrict the comparison to S/N $>$ 20 to provide a comparison with \textit{Gaia} on the same scale as the other samples. Both DB samples already exclude magnetic white dwarfs, and therefore we make no additional modifications. We also make no distinction between the SDSS spectrographs in this case because the number of objects is much smaller (see Table~\ref{WDsample}) and the two adopted samples use different models and fitting techniques, meaning that isolating data reduction issues is much more difficult.

\begin{figure}
\centering
\includegraphics[width=0.55\columnwidth,bb = 85 250 400 520]{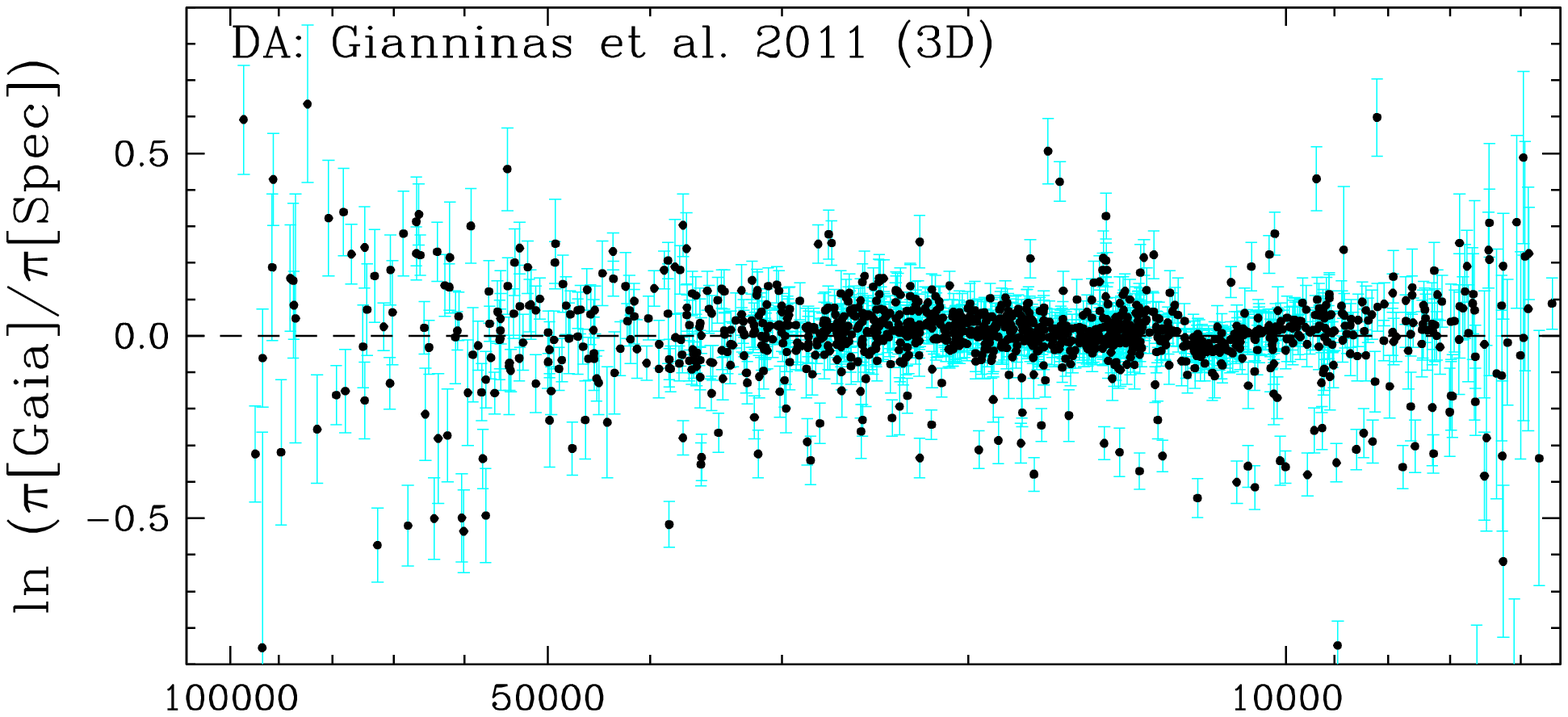}
\newline
\includegraphics[width=0.55\columnwidth,bb = 85 250 400 520]{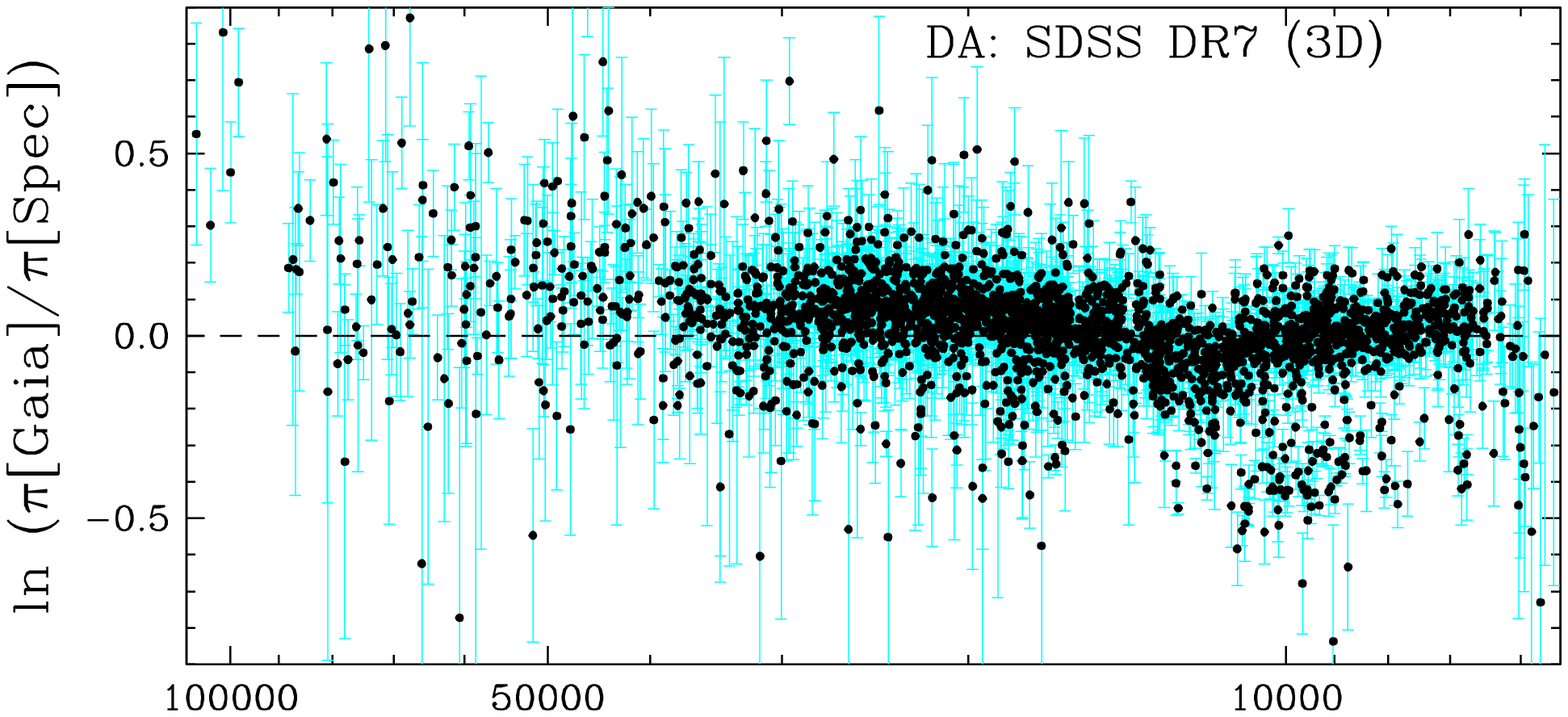}
\newline
\includegraphics[width=0.55\columnwidth,bb = 85 250 400 520]{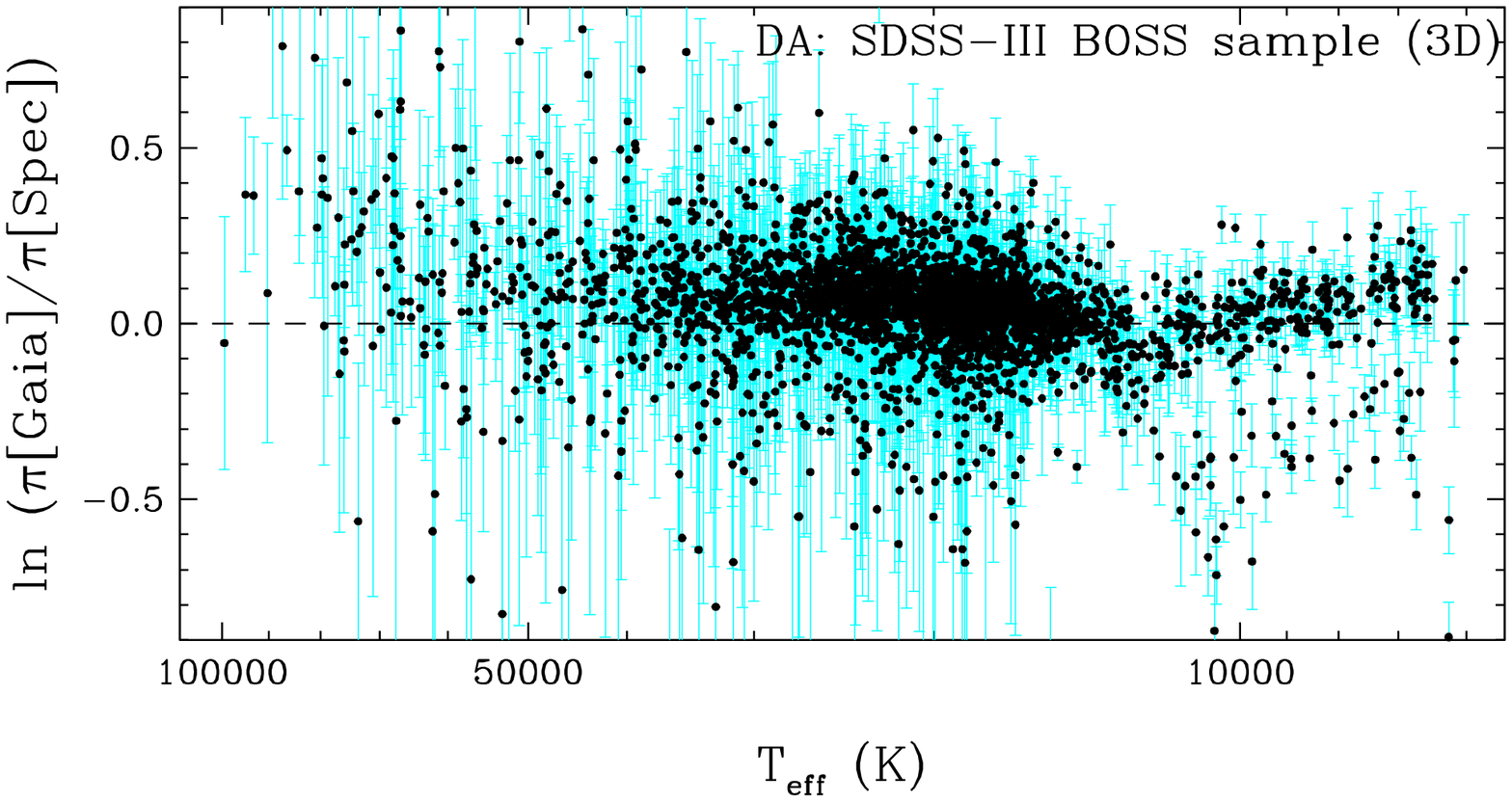}
\newline
\caption{Natural logarithm of the ratio between observed \textit{Gaia} DR2 and predicted spectroscopic parallaxes using 3D corrections for the samples of \citet{gianninas11} (top panel), SDSS DR7 (middle panel), and SDSS-III BOSS (bottom panel). We note that the natural logarithm of the ratio is close to the difference in percentage.\label{fig1}}
\end{figure}

We have cross-matched our initial samples with the \textit{Gaia} DR2 white dwarf catalogue of \citet{gentilefusillo18}. A number of \textit{Gaia} sources have poor quality flags, and we made the following additional cuts
\begin{align}
\textsc{astrometric\_excess\_noise} < 0.75\label{piercut3}~,\\
\textsc{phot\_bp\_rp\_excess\_factor}/\nonumber\\
  (1.3+0.06(G_{\rm BP}-G_{\rm RP})^2) < 1.0~,\\
\textsc{astrometric\_sigma5d\_max} < 1.0\label{piercut4}~.
\end{align}
{\noindent}The removal of a small fraction of lower quality data with likely underestimated \textit{Gaia} errors does not significantly impact our analysis. We note that these cuts, coupled with the removal of DA+DC and DA+DB candidates based on a spectral visual inspection, implies that we have biases against detecting double degenerates in our adopted samples. Therefore, we make no attempt to derive the double degenerate fraction. The different samples are nevertheless a reasonable representation of the magnitude-limited local population of non-magnetic and apparently single white dwarfs, and completeness is not required to compare their fundamental parameters using different methods. The final number of white dwarfs with precise spectroscopic atmospheric parameters and clean \textit{Gaia} data is given in Table~\ref{WDsample}. 

The adopted samples are rather different in terms of the volume covered, with average parallaxes of 15.4, 6.1, and 3.8 mas for \citet{gianninas11}, SDSS DR7, and SDSS-III BOSS, respectively. In the latter case, the upgraded and bluer sensitive BOSS spectrograph is allowing the observation of more distant white dwarfs at a higher S/N compared to what had been possible before. Given that \textit{Gaia} precision decreases with both magnitude and distance, it suggests that the SDSS-III BOSS sample will provide a less thorough test of the consistency between \textit{Gaia} and spectroscopic analyses.

We use the dereddened \textit{Gaia} $G$, $G_{\rm BP}$ and $G_{\rm RP}$ magnitudes, parallaxes, and derived photometric atmospheric parameters as made available in \citet{gentilefusillo18}. In brief, the magnitudes are dereddened using the 2D maps (total line-of-sight reddening) of \citet{schlegel98} with corrections from \citet{schlafly11}. For the third spatial dimension, the distance, we use the reddening law described in equations 17-19 of \citet{gentilefusillo18} where a gas scale height of 200\,pc was constrained empirically from the \textit{Gaia} white dwarf cooling sequence. The dereddened \textit{Gaia} magnitudes are then converted to observed fluxes $f_\lambda^S$ in erg cm$^{-2}$ s$^{-1}$ in the revised \textit{Gaia} DR2 passbands $S_{\lambda}$ \citep{gaiaDR2-ArXiV-3} using the appropriate zero points \citep[see table 3 of][]{gentilefusillo18}. The minimised quantity is $f_\lambda^S = 4 \pi \varpi^2 R^2 H^{S}_\lambda(T_{\rm eff},\log g)$ where $R$ is the white dwarf radius, $\varpi$ is the parallax in arcsec, and $H^{S}_\lambda$ is the Eddington flux predicted from model atmospheres integrated over the \textit{Gaia} passbands. Only $T_{\rm eff}$ and $\log g$ are kept as free parameters and the radius is fixed from the mass-radius relation of \citet{fontaine01} for C/O-cores (50/50 by mass fraction mixed uniformly). We use thick hydrogen layers ($M_{\rm H}/M_{\rm WD} = 10^{-4}$) and thin hydrogen layers ($M_{\rm H}/M_{\rm WD} = 10^{-10}$) for DA and DB white dwarfs, respectively. For $T_{\rm eff} >$ 30\,000~K, we interpolate over the pure C-core sequences of \citet{wood95}. For masses below 0.46~$M_{\odot}$, we use the He-core cooling sequences of \citet{althaus01}. 

Before comparing the photometric and spectroscopic analyses in two dimensions ($T_{\rm eff}$ and $\log g$) in Section~\ref{sec4}, we proceed with an intermediate step in Section~\ref{sec3} where we neglect \textit{Gaia} $G_{\rm BP}$ and $G_{\rm RP}$ in order to compare both techniques in 1D dimension, here with the independent variable chosen to be the parallax. In doing so, we use the same models as described above, but do not fit \textit{Gaia} photometry and instead use the spectroscopic atmospheric parameters coupled with predicted stellar fluxes and the \textit{Gaia} $G$ magnitude to predict a {\it spectroscopic parallax}. This makes any potential problem caused by the \textit{Gaia} photometric calibration more easily tractable. 

\section{Predicted and Observed Parallaxes}
\label{sec3}

\begin{figure}
\centering
\includegraphics[width=0.55\columnwidth,bb = 85 250 400 520]{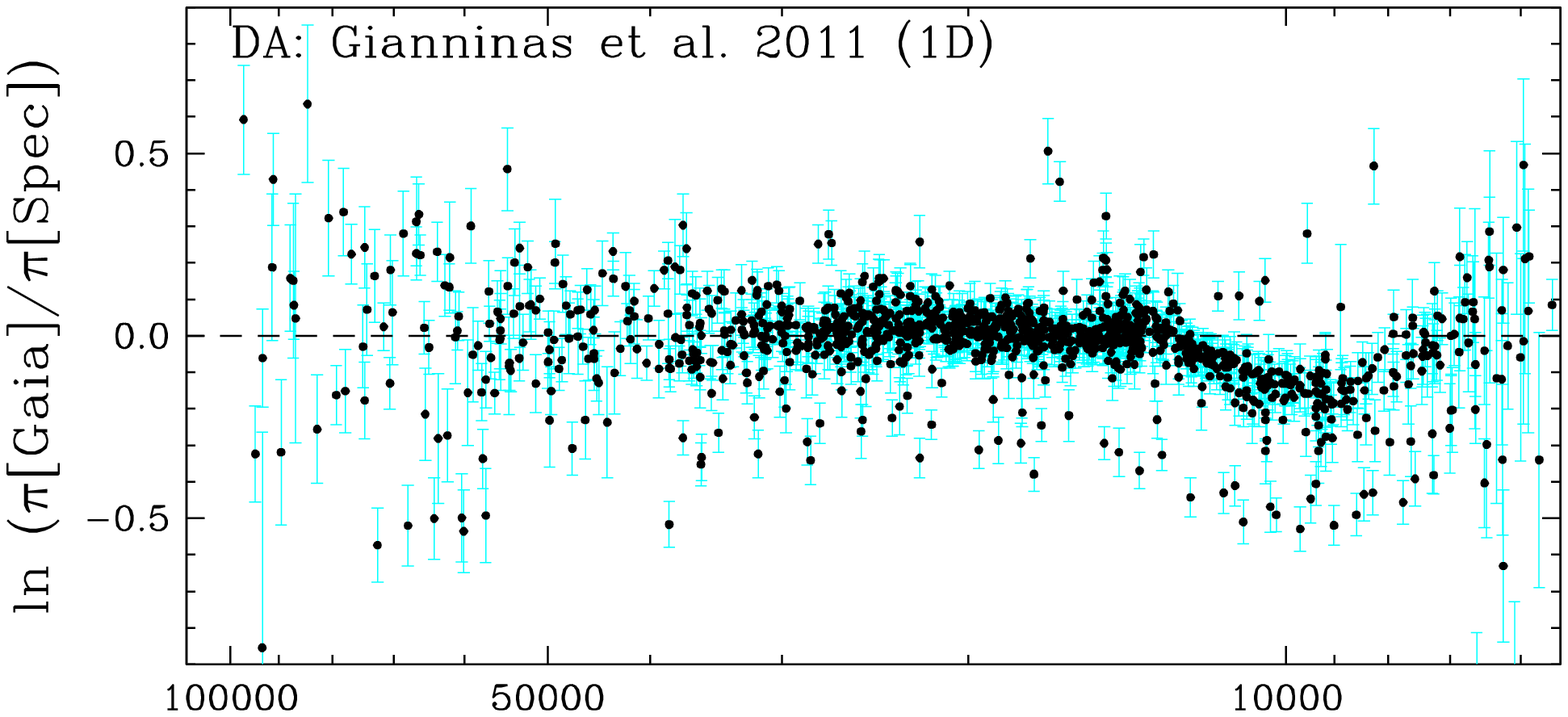}
\newline
\includegraphics[width=0.55\columnwidth,bb = 85 250 400 520]{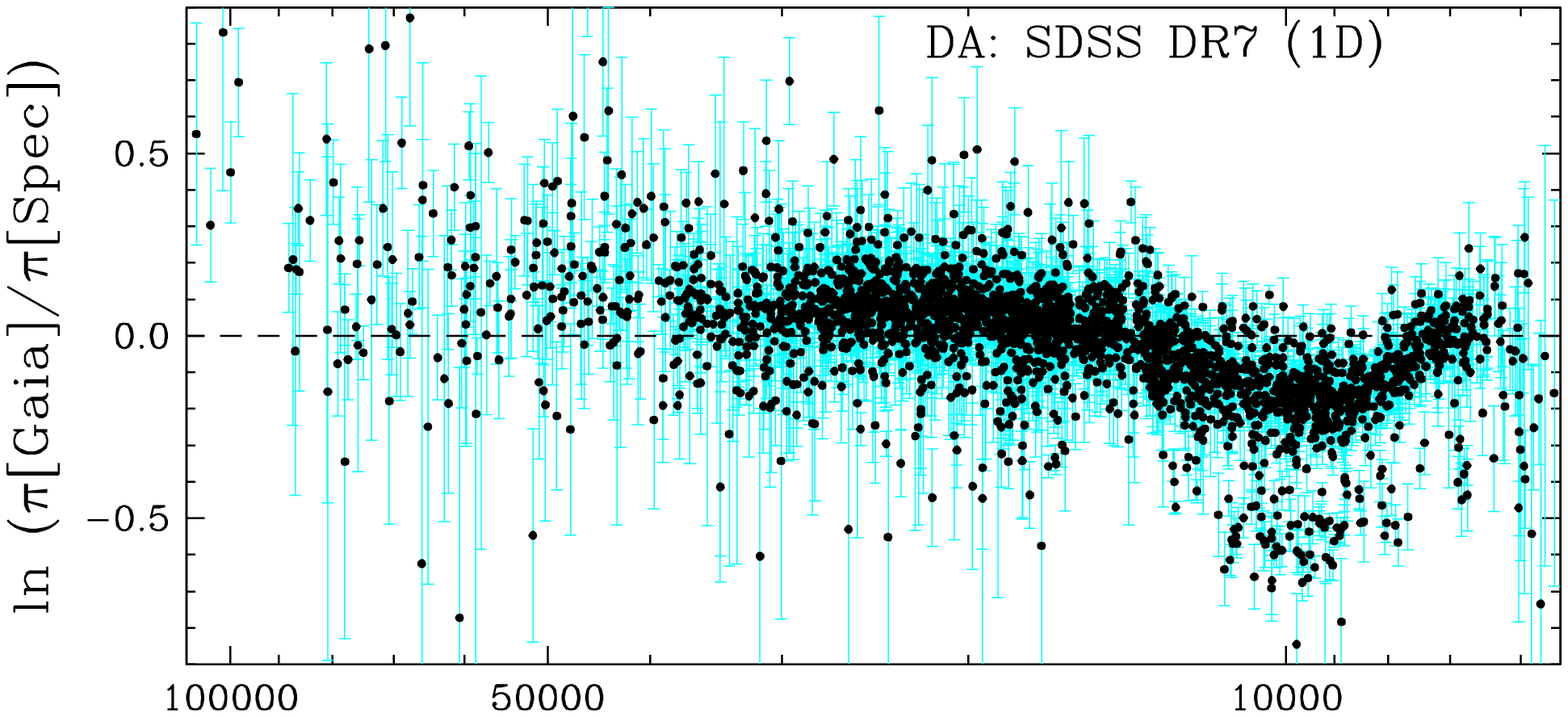}
\newline
\includegraphics[width=0.55\columnwidth,bb = 85 250 400 520]{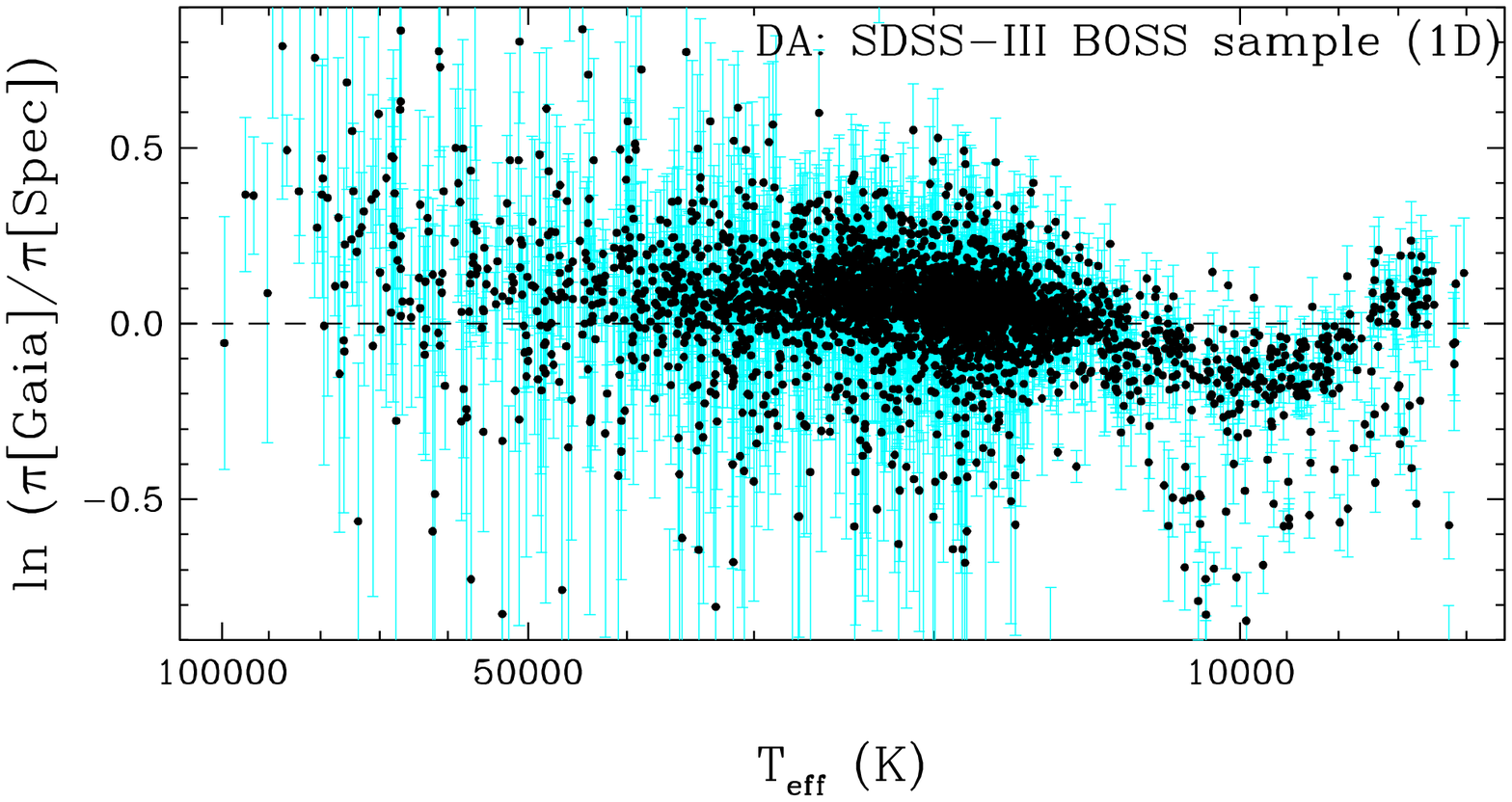}
\newline
\caption{Natural logarithm of the ratio between observed \textit{Gaia} DR2 and predicted spectroscopic parallaxes for the same samples as in Fig.~\ref{fig1} but with 1D model atmospheres. \label{fig2}}
\end{figure}

\begin{figure}
\centering
\includegraphics[width=0.55\columnwidth,bb = 125 220 440 520]{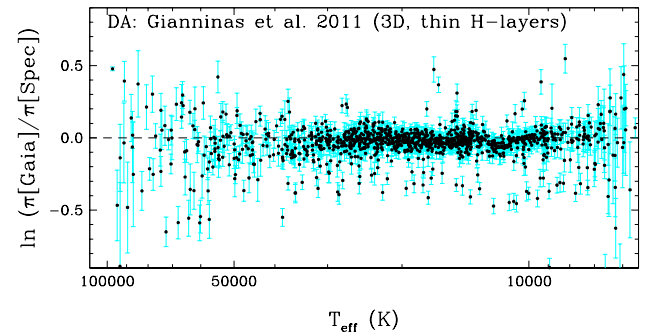}
\caption{Natural logarithm of the ratio between observed \textit{Gaia} DR2 and predicted spectroscopic parallaxes using 3D corrections and assuming thin H-layers for the \citet{gianninas11} sample. \label{fnew}}
\end{figure}

The comparison of \textit{Gaia} DR2 parallaxes with predicted values from spectroscopic analyses of DA white dwarfs is presented in Fig.~\ref{fig1} when the atmospheric parameters are corrected for 3D convective effects \citep{tremblay13}. The average spread clearly increases going from bright \citep{gianninas11} to faint white dwarfs (SDSS-III BOSS), which is mainly caused by increasing \textit{Gaia} error bars. Furthermore, the SDSS-III BOSS survey has fewer cool white dwarfs ($T_{\rm eff} < 12\,000$~K) owing to a change in the spectroscopic follow-up selection function. Nevertheless there are enough objects over all selected samples to look at systematic trends. 

In all three samples, an inflexion is seen in the range between 11\,000~K and 13\,000~K, where the spectroscopic parallaxes are slightly under luminous. While this is coincident with a high-mass problem \citep{tremblay10}, it is also close to the location of the maximum strength of the Balmer lines where $T_{\rm eff}$ determinations are less precise \citep{bergeron95}. Fig.~\ref{fig2} demonstrates that when using 1D model atmospheres, there is a strong discrepancy with \textit{Gaia} for all three samples. This confirms that the 3D model atmospheres, including improved physics, are in much better agreement with the observations \citep{tremblay13}. In the range 11\,000-13\,000~K, it is tempting to suggest that there is a small, residual high-mass problem according to Fig.~\ref{fig1}. This is further investigated in Section~\ref{sec4} making use of photometric parameters.

At warm $T_{\rm eff}$ values above 13\,000~K where convective flux is negligible, we do not observe a systematic, cross-survey offset between spectroscopic and trigonometric parallaxes. However, both SDSS samples suggest a trend of luminosities that are over-predicted  at high temperatures, which is not seen in the \citet{gianninas11} sample. One reason could be an issue with our approximate treatment of dereddening \citep{gentilefusillo18}, but this is likely ruled out because the SDSS-III BOSS sample, which covers a larger volume, is not in significantly worse agreement with \textit{Gaia}. Furthermore, all data sets were analysed with the same models and fitting technique. One explanation for the behaviour is within the spectroscopic calibration of the SDSS spectrograph. It has been reported on many occasions \citep{kleinman04,tremblay11,tremblay16} that SDSS spectroscopic parameters are systematically offset from those of independent surveys but the issue remains unchanged with the DR14 data reduction. However, the SDSS-III BOSS spectrograph with the DR14 data reduction appears to show the same trend. Since both SDSS spectrographs share a data reduction pipeline, it may be that the two SDSS samples are not fully independent and one should be cautious in the interpretation of the differences with the \citet{gianninas11} sample.

Fig.~\ref{fnew} demonstrates that if thin instead of thick hydrogen layers are employed for the bright DA stars in \citet{gianninas11}, the agreement is only marginally worse between spectroscopic and trigonometric parallaxes. It confirms that individual uncertainties are too large to assign hydrogen layer thicknesses on a case-by-case basis \citep{tremblay17,joyce18}. Additionally there is no clear split in the \citet{gianninas11} distribution of Fig.~\ref{fig1} that would suggest a bimodal distribution of H-layer thicknesses. Whether any information on the mass-radius relation or hydrogen layer masses as a function of white dwarf mass \citep{romero12} can be extracted from these data sets will need to be considered from a more careful statistical analysis with a better understanding of the spectroscopic error bars.

There are a number of outliers in all surveys for which we have carefully examined the data. A significant number of these objects could be DA+DA or DA+DC unresolved double degenerates for which the optical spectrum behaves like a single star. In particular, Figs.~\ref{fig1}-\ref{fig2} confirm a fairly obvious sequence of objects at cool $T_{\rm eff}$ below the standard white dwarf sequence where the \textit{Gaia} fluxes are over-luminous by a factor of about of two, correspond to a parallax offset of a factor 1.5 ($\ln[\pi_{\rm Gaia}/\pi_{\rm Spectro}] \approx -0.35$). These are likely to be unresolved DA+DA double degenerates. DA+DC double degenerates have much more varied spectroscopic solutions owing to the arbitrary dilution of the Balmer lines \citep{tremblay11}, and are suspected to be the source of some of the outliers in other parts of the diagram. In some cases, these objects could also be single, helium-rich DA white dwarfs \citep{gentile17}. Finally, a fraction of outliers are caused by a confusion between the cool and hot spectroscopic solutions \citep{bergeron92}, spectroscopic flux calibration issues, or \textit{Gaia} data issues without raised quality flags. While outside of the scope of this work, this demonstrates the potential of this technique to unravel statistics on the population of double degenerates and helium-rich DA white dwarfs. We note that there are more outliers in the \citet{gianninas11} sample than in the SDSS, which is likely because the latter data set includes $ugriz$ photometry and a spectroscopic coverage up to $\approx$ 1 $\mu m$, which allowed us to more easily remove binaries and select between the cool and hot solutions at the pre-analysis stage.

\begin{figure}
\centering
\includegraphics[width=0.55\columnwidth,bb = 85 250 400 520]{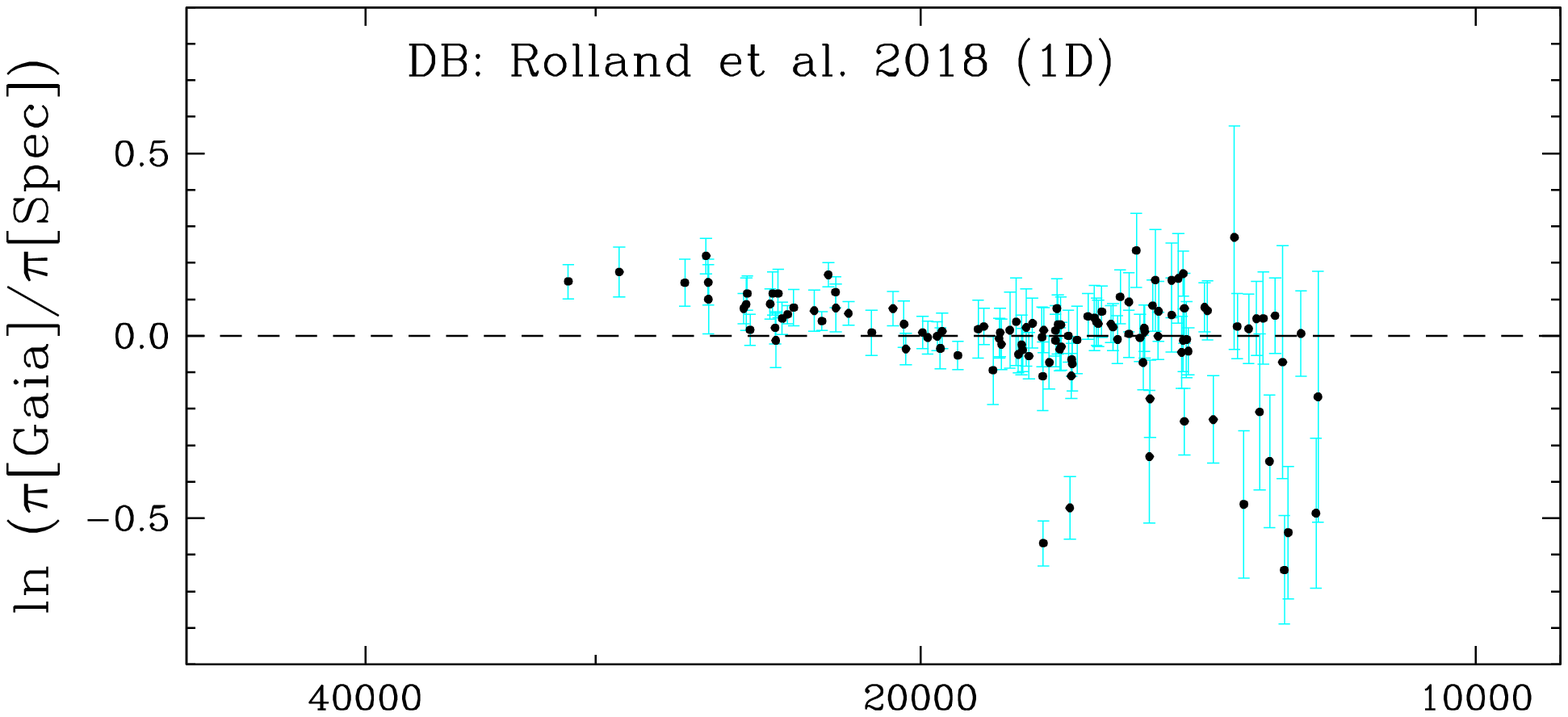}
\newline
\includegraphics[width=0.55\columnwidth,bb = 85 250 400 520]{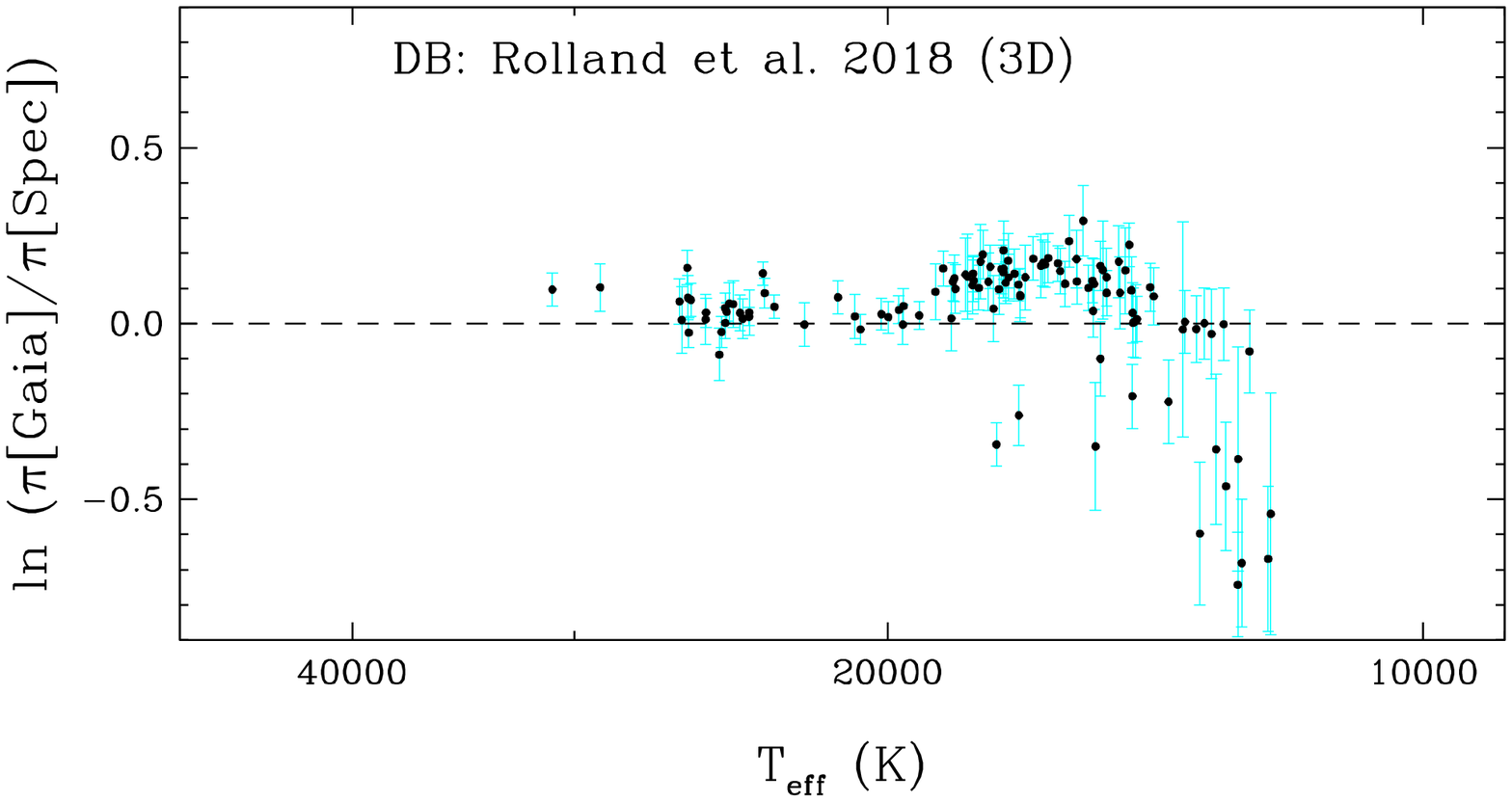}
\newline
\caption{Natural logarithm of the ratio between observed \textit{Gaia} DR2 and predicted spectroscopic parallaxes for the DB/DBA sample of \citet{rolland18} using 1D model atmospheres \citep[][top panel]{bergeron11} and with 3D corrections \citep[][bottom panel]{cukanovaite18}. \label{fig3}}
\end{figure}

The comparison of the spectroscopic analysis of DB and DBA white dwarfs from \citet{rolland18} with \textit{Gaia} data is shown in Fig.~\ref{fig3} (upper panel). All these objects have a convective atmosphere and \citet{cukanovaite18} have recently proposed 3D corrections assuming pure-He composition. Fig.~\ref{fig3} (bottom panel) illustrates the effects of 3D modeling, with the important caveat that this is a preliminary assessment without accounting for the effect of hydrogen on 3D corrections, as the majority of DB white dwarfs are thought to have traces of hydrogen \citep{koester15,rolland18}. The parameters of DB stars are in reasonable agreement with \textit{Gaia}, although for cool objects ($T_{\rm eff} < 14\,000~$K) where a problem with neutral line broadening is suspected \citep{rolland18}, \textit{Gaia} clearly suggests lower masses than predicted from spectroscopy. 
 
\begin{figure}
\centering
\includegraphics[width=0.55\columnwidth,bb = 85 250 400 520]{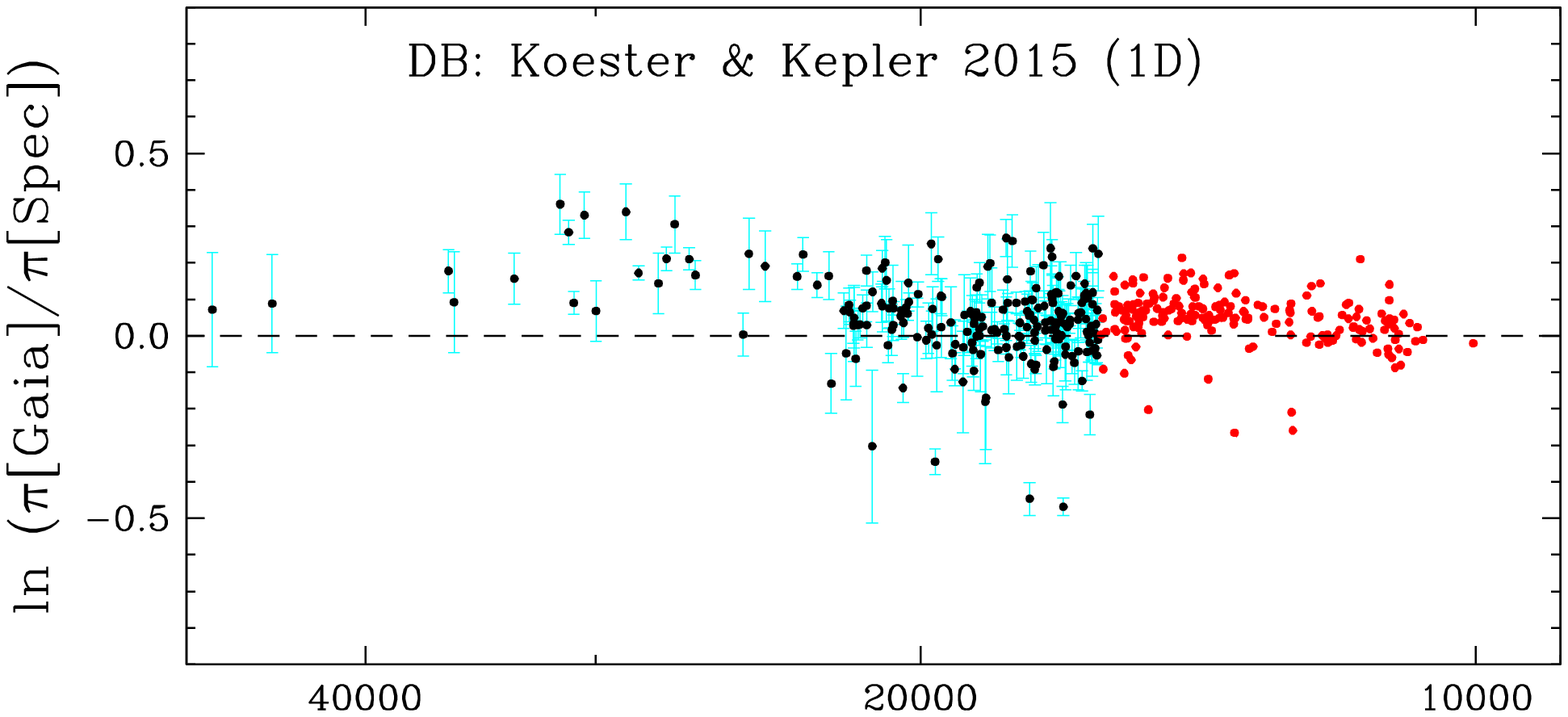}
\newline
\includegraphics[width=0.55\columnwidth,bb = 85 250 400 520]{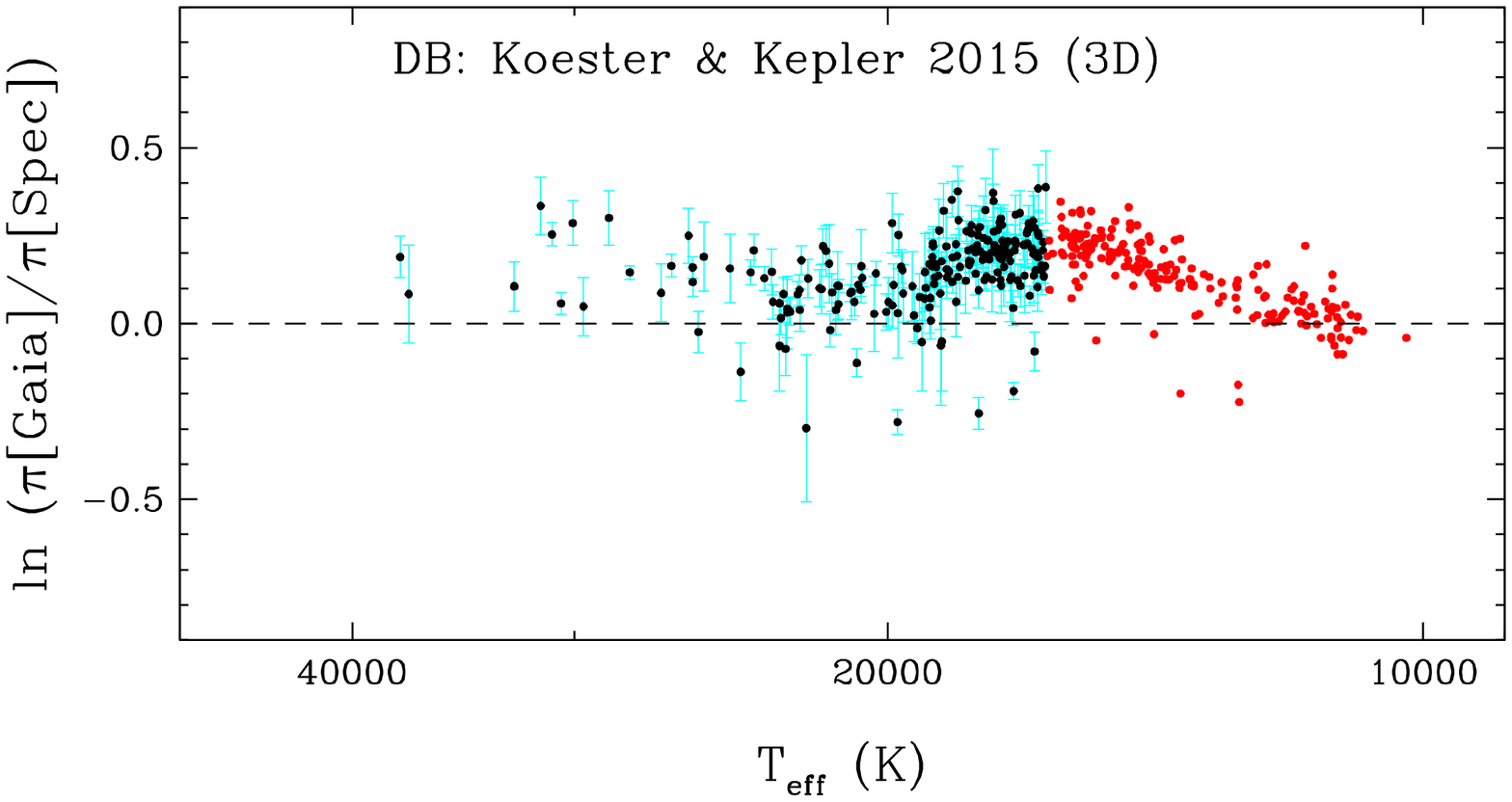}
\newline
\caption{Same as Fig.~\ref{fig3} but for the SDSS DB/DBA sample and model atmosphere fits of \citet{koester15}. Objects with fixed spectroscopic parameters at a value of $\log g = 8.0$ are plotted as red points with no error bars.  \label{fig4}}
\end{figure}

We also present the comparison of \textit{Gaia} with the SDSS DB and DBA spectroscopic sample of \citet{koester15} in Fig.~\ref{fig4}, once again both for 1D and 3D model atmospheres. For objects below $T_{\rm eff} = 16\,000$~K, \citet{koester15} have assumed $\log g = 8.0$ due to potential issues with neutral line broadening. We still show these stars in Fig.~\ref{fig4} but with red points and no error bars. The two published analyses of DB white dwarfs presented in this paper use different models and fitting methods but the agreement with \textit{Gaia} is similar according to Figs.~\ref{fig3} and \ref{fig4}. For $T_{\rm eff} \gtrapprox 16\,000$~K, both the 3D and 1D atmospheric parameters agree equally well with \textit{Gaia} within 2$\sigma$ on a star-by-star basis.

The parameters of cool SDSS DB stars where $\log g = 8.0$ was assumed do agree with \textit{Gaia} parallaxes with a very small scatter. This suggests that the mass distribution of DB stars does not have a high-mass problem when the photometric technique is employed, and that the discrepancy is entirely caused by issues with the spectroscopic technique.

\section{The Temperature and Surface Gravity Scale}
\label{sec4}

The predicted spectroscopic parallax depends on the absolute magnitude which is itself a function of both $T_{\rm eff}$ and $\log g$. As a consequence, it is of interest to study whether both spectroscopic parameters do agree with independent \textit{Gaia} measurements. 
 
We directly employ the photometric atmospheric parameters derived in section~4 of \citet{gentilefusillo18} and briefly described earlier in Section~\ref{sec2}. \citet{gentilefusillo18} have demonstrated that \textit{Gaia} photometric fits are in very good agreement with fits using instead Pan-STARRS DR1 or SDSS $ugriz$ photometry. As a starting point for this study, we assume that \textit{Gaia} photometric parameters are as accurate as the model atmospheres, mass-radius relation, and dereddening procedure can describe them. We note that $T_{\rm eff}$ almost only depends on \textit{Gaia} colours and is degenerate with the fixed dereddening procedure. On the other hand, $\log g$ almost only depends on the absolute \textit{Gaia} fluxes and is degenerate with the fixed mass-radius relation. 

\begin{figure}
\centering
\includegraphics[width=0.55\columnwidth,bb = 85 150 400 590]{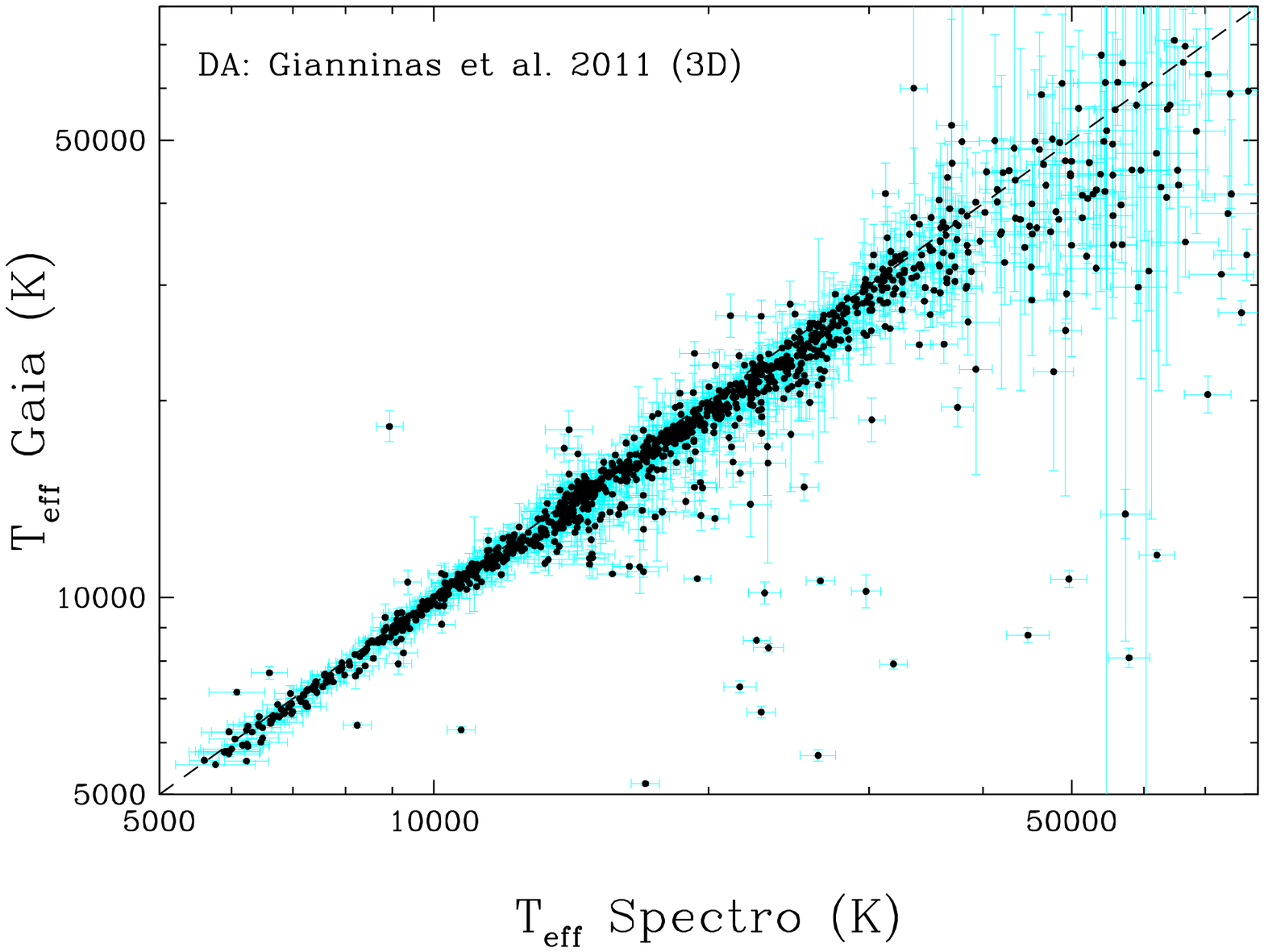}
\newline
\includegraphics[width=0.55\columnwidth,bb = 85 150 400 590]{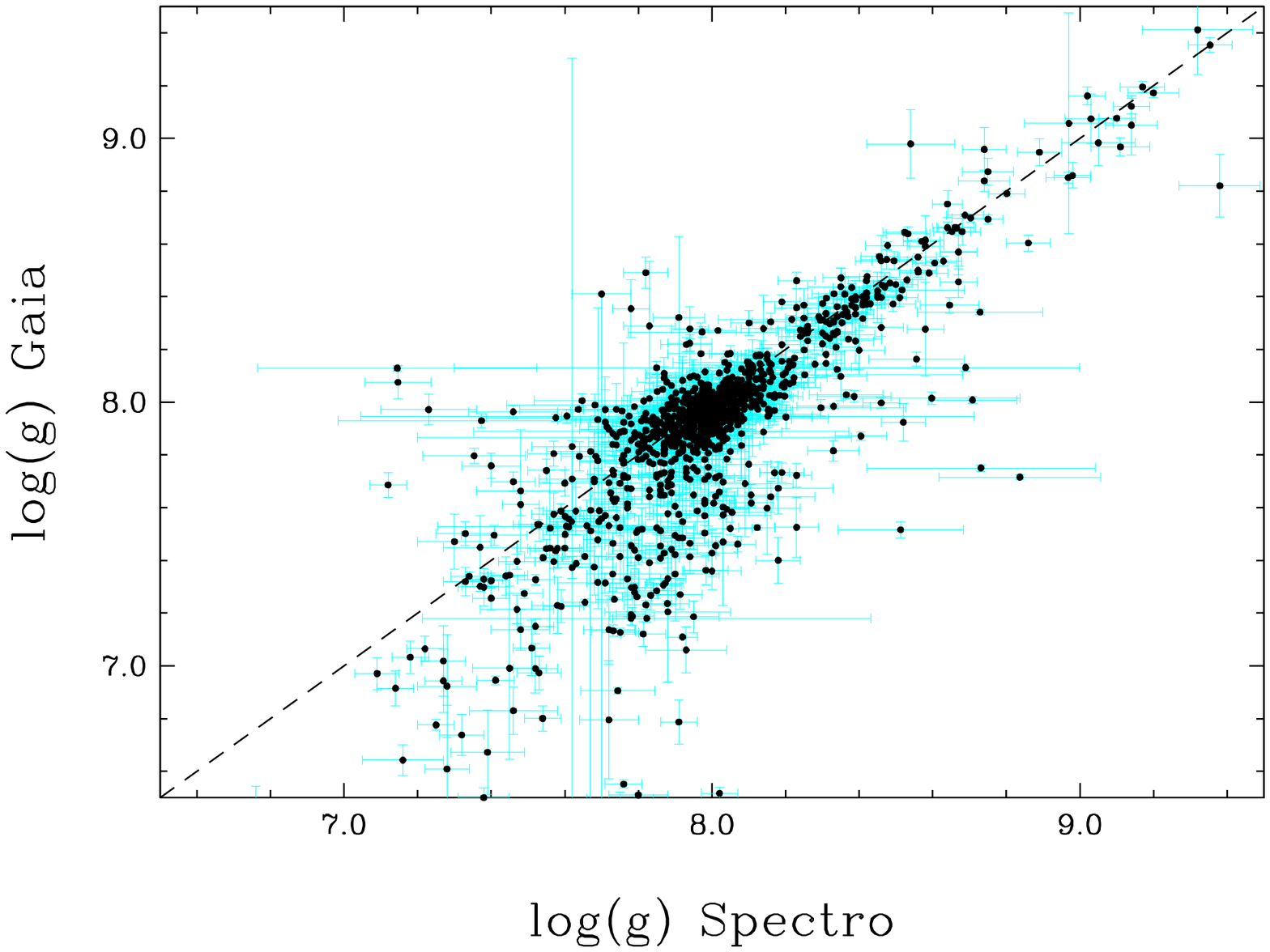}
\newline
\caption{Comparison of spectroscopic and photometric \textit{Gaia} effective temperatures (top panel) and surface gravities (bottom panel) 
for the DA sample of \citet{gianninas11} corrected for 3D effects \citep{tremblay13}. The one-to-one agreement is illustrated by the dashed lines. \label{fig5}}
\end{figure}

\begin{figure}
\centering
\includegraphics[width=0.55\columnwidth,bb = 85 150 400 590]{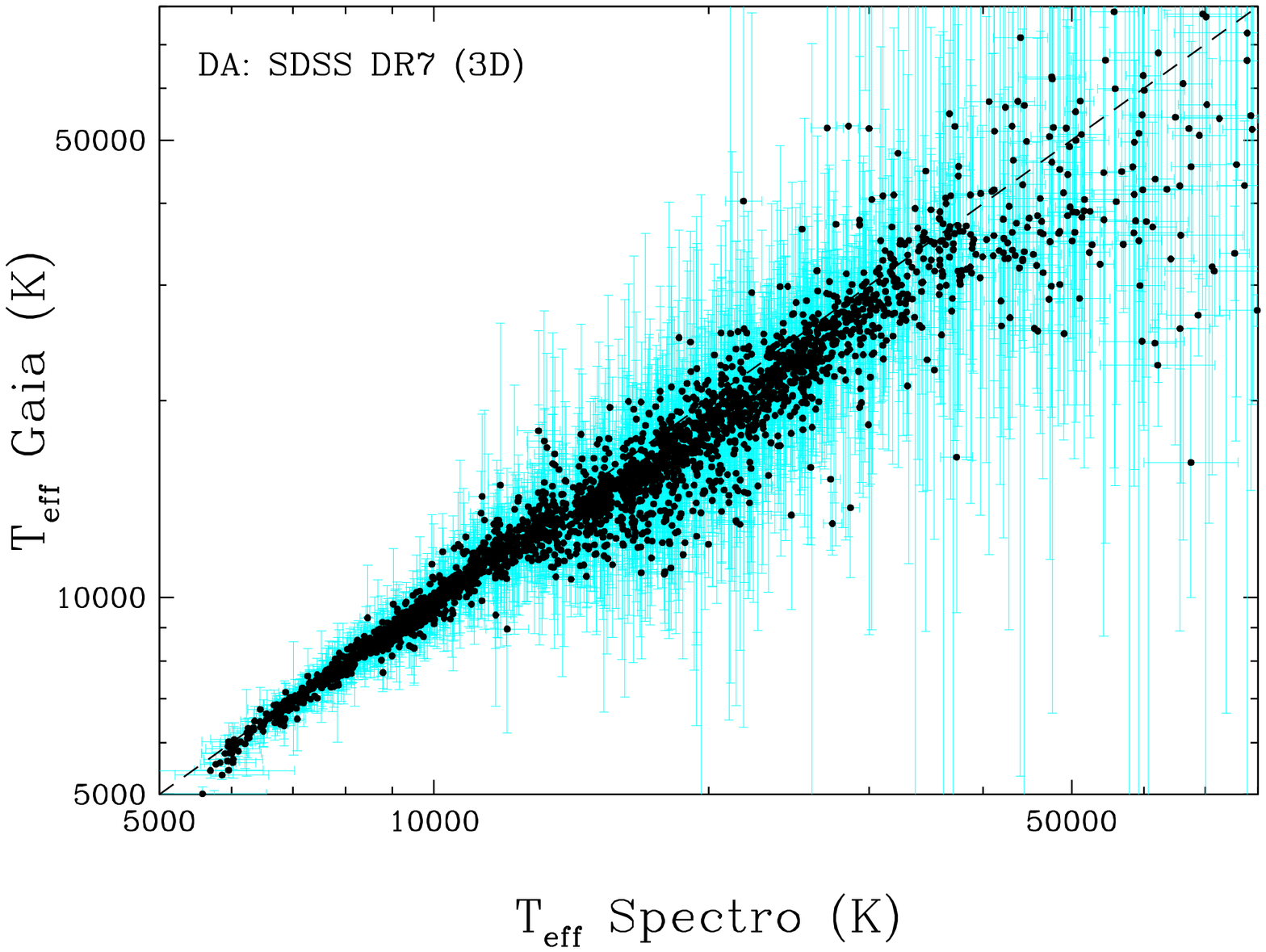}
\newline
\includegraphics[width=0.55\columnwidth,bb = 85 150 400 590]{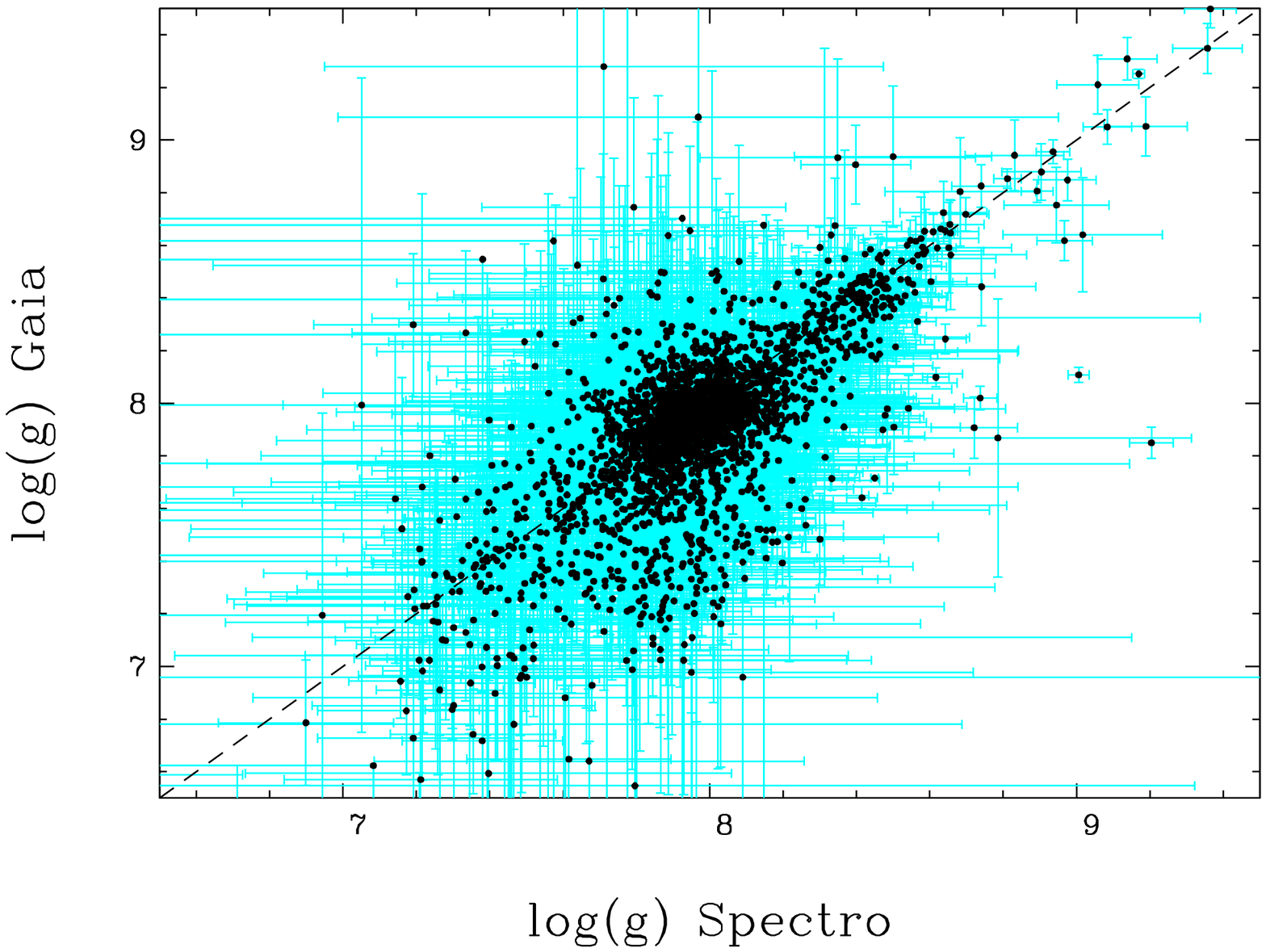}
\newline
\caption{Same as Fig.~\ref{fig5} but for the SDSS DR7 sample. \label{fig6}}
\end{figure}

\begin{figure}
\centering
\includegraphics[width=0.55\columnwidth,bb = 85 150 400 590]{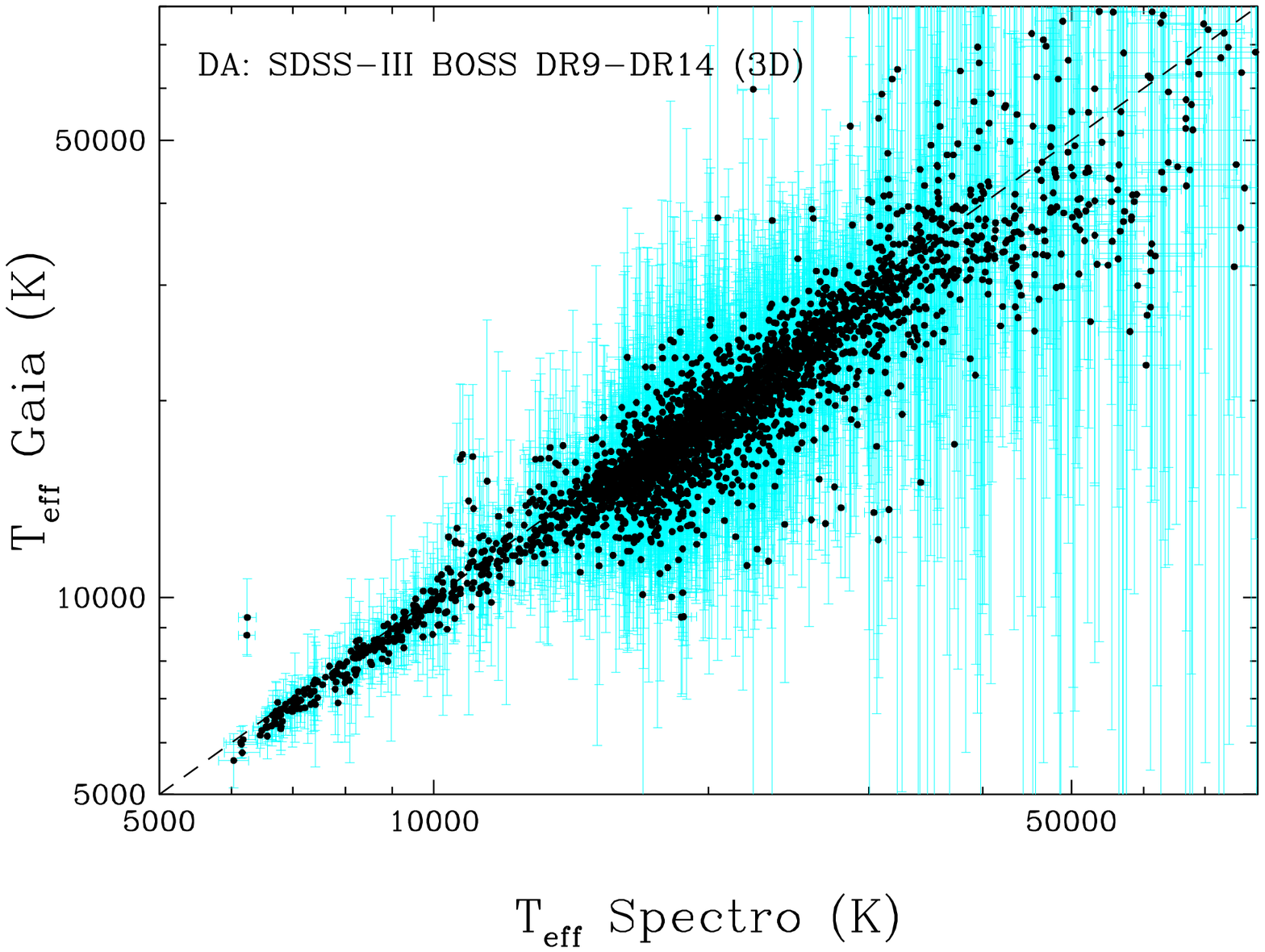}
\newline
\includegraphics[width=0.55\columnwidth,bb = 85 150 400 590]{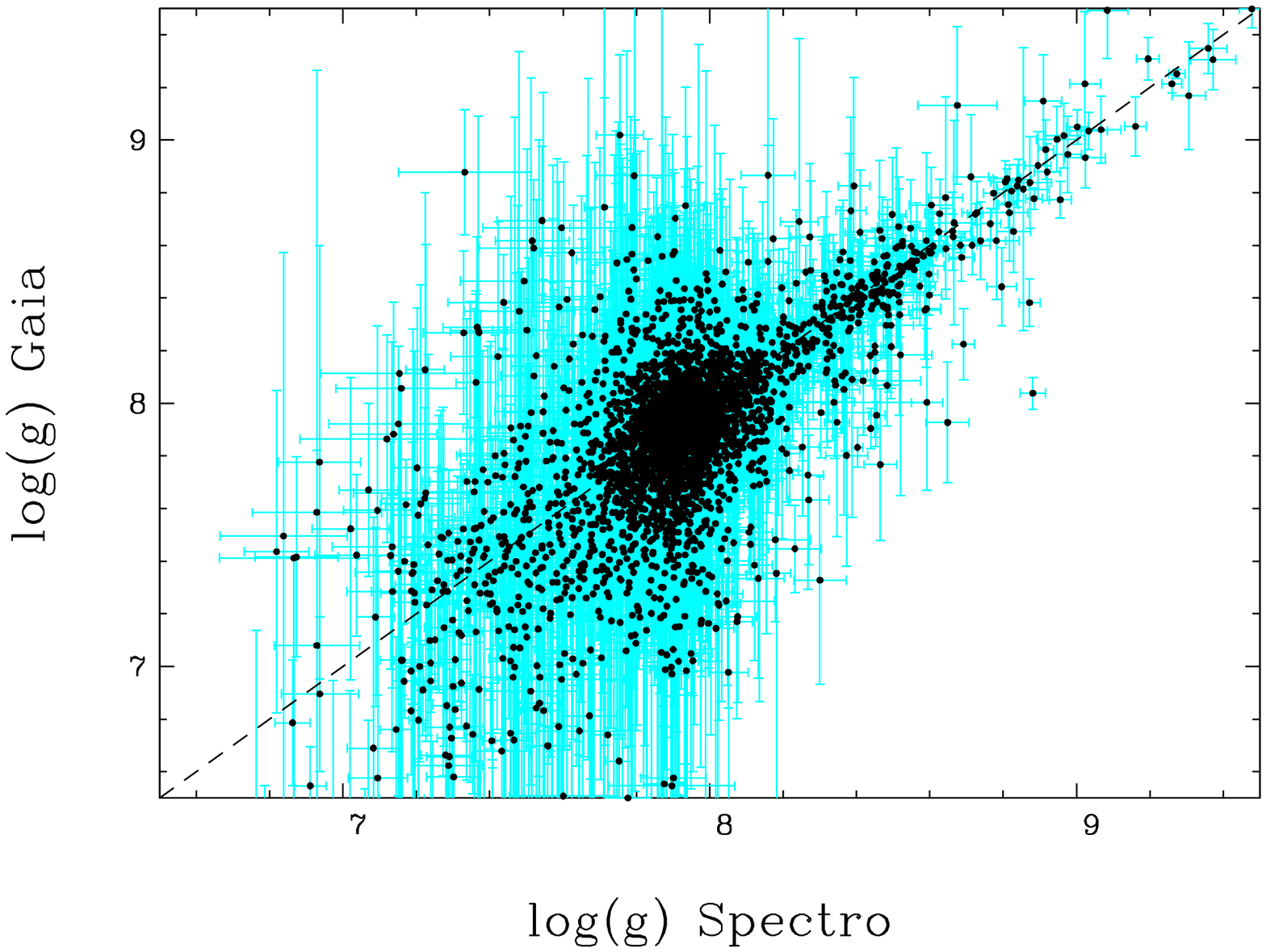}
\newline
\caption{Same as Fig.~\ref{fig5} but for the SDSS-III BOSS sample. \label{fig7}}
\end{figure}

Figs.~\ref{fig5}-\ref{fig7} present the comparison of photometric and 3D spectroscopic parameters for our samples of DA white dwarfs. In all cases we can see similar trends, with the \textit{Gaia} effective temperatures being systematically smaller than the spectroscopic values for all $T_{\rm eff}$, even though individual stars are in agreement within 1-2$\sigma$. The offset was not observed in Section~\ref{sec3} where photometric $T_{\rm eff}$ values were not employed, which could be suggesting an issue with reddening. However, the fact that similar trends are seen for all three samples covering very different volumes suggests that reddening is not the obvious culprit, and we have verified that unrealistic large reddening, i.e. close to the full line-of-sight independently of the distance, is needed to make the temperature scales in agreement. We conclude that a slight issue with the spectroscopic temperature scale is a more likely explanation. The discrepancy is slightly more important for the SDSS spectrograph, suggesting once again an issue with the flux calibration for observations taken up to DR7. The comparison of $\log g$ values has less obvious trends since most objects concentrate around $\log g = 8.0$, although there is a slight systematic offset that largely cancels out with the temperature offset when calculating absolute magnitudes, which explains the good agreement in Section~\ref{sec3}. It is interesting to remark that when non-ideal effects were included in currently employed Stark broadening tables \citep{tremblay09}, both $\log g$ and $T_{\rm eff}$ changed in the same direction so that the predicted absolute magnitude remained largely constant. Therefore, it is tantalizing to suggest that \textit{Gaia} may be observing some residual issues in the line broadening. A similar result was found in \citet{genest14} where they used SDSS photometry. 

\begin{figure}
\centering
\includegraphics[width=0.55\columnwidth,bb = 85 250 400 520]{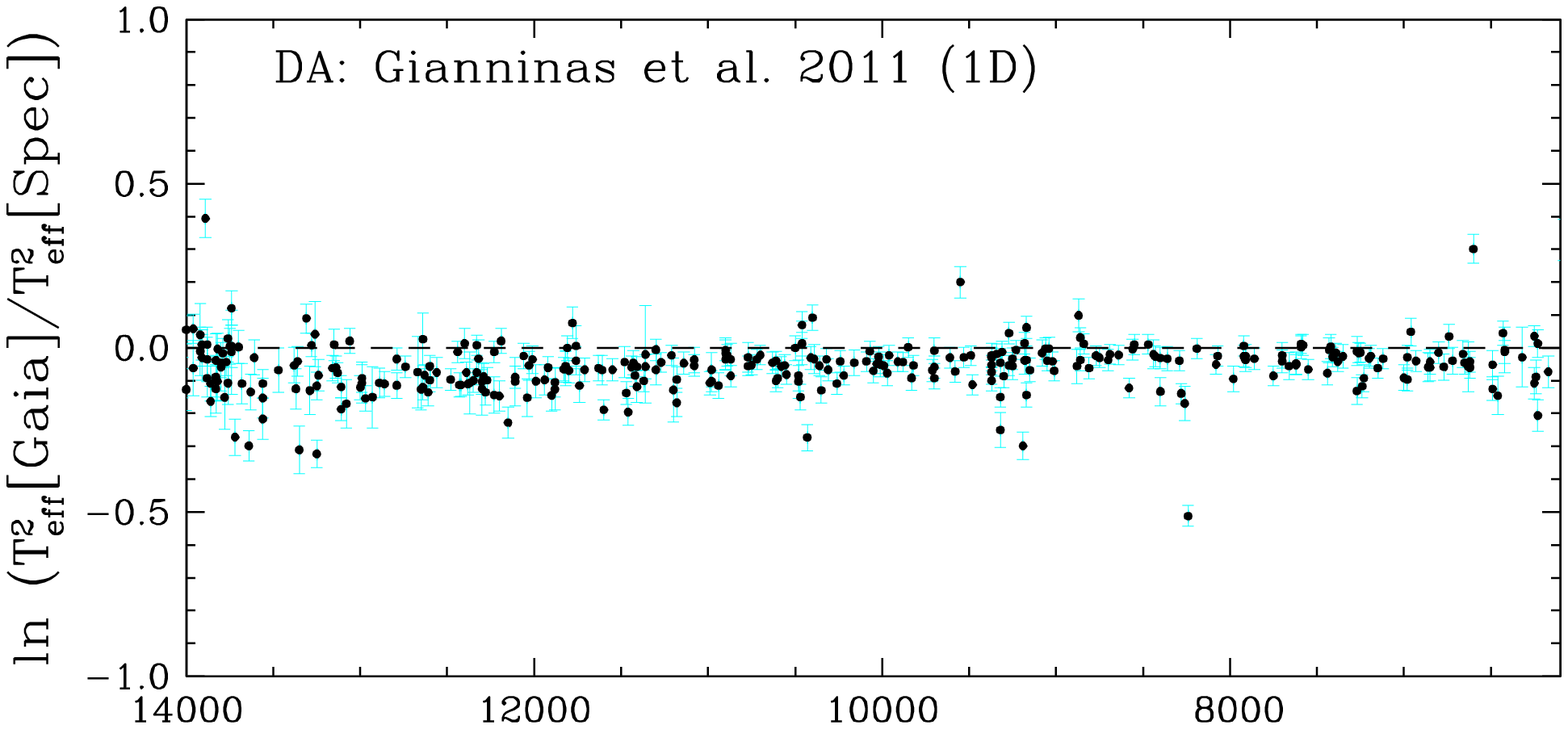}
\newline
\includegraphics[width=0.55\columnwidth,bb = 85 250 400 520]{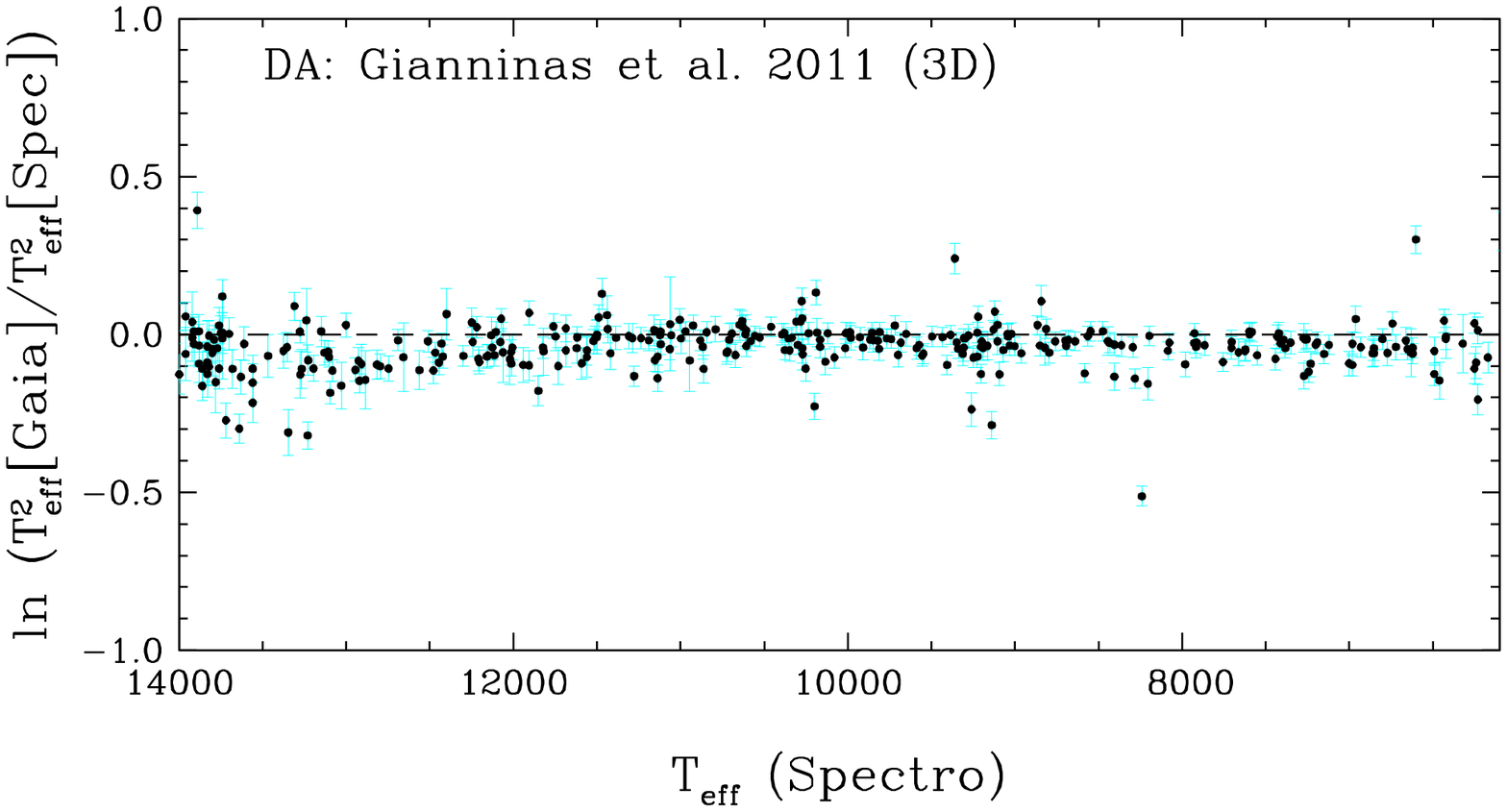}
\newline
\caption{Natural logarithm of the $T_{\rm eff}^2$ ratio between Gaia photometric fits and \citet{gianninas11} spectroscopic parameters as a function of effective temperature for both 1D (top panel) and 3D model atmospheres (bottom panel). \label{fig8}}
\end{figure}

\begin{figure}
\centering
\includegraphics[width=0.55\columnwidth,bb = 85 250 400 520]{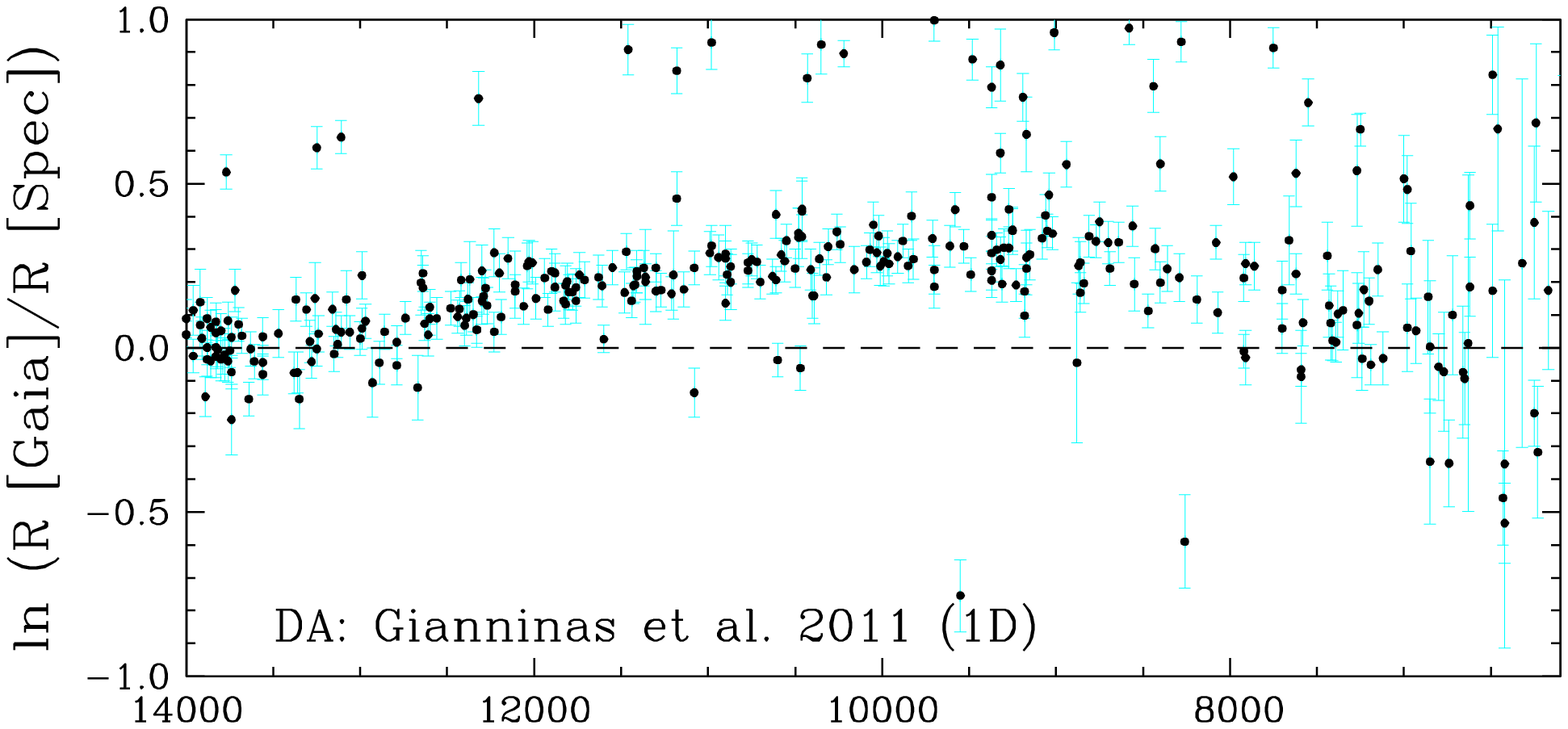}
\newline
\includegraphics[width=0.55\columnwidth,bb = 85 250 400 520]{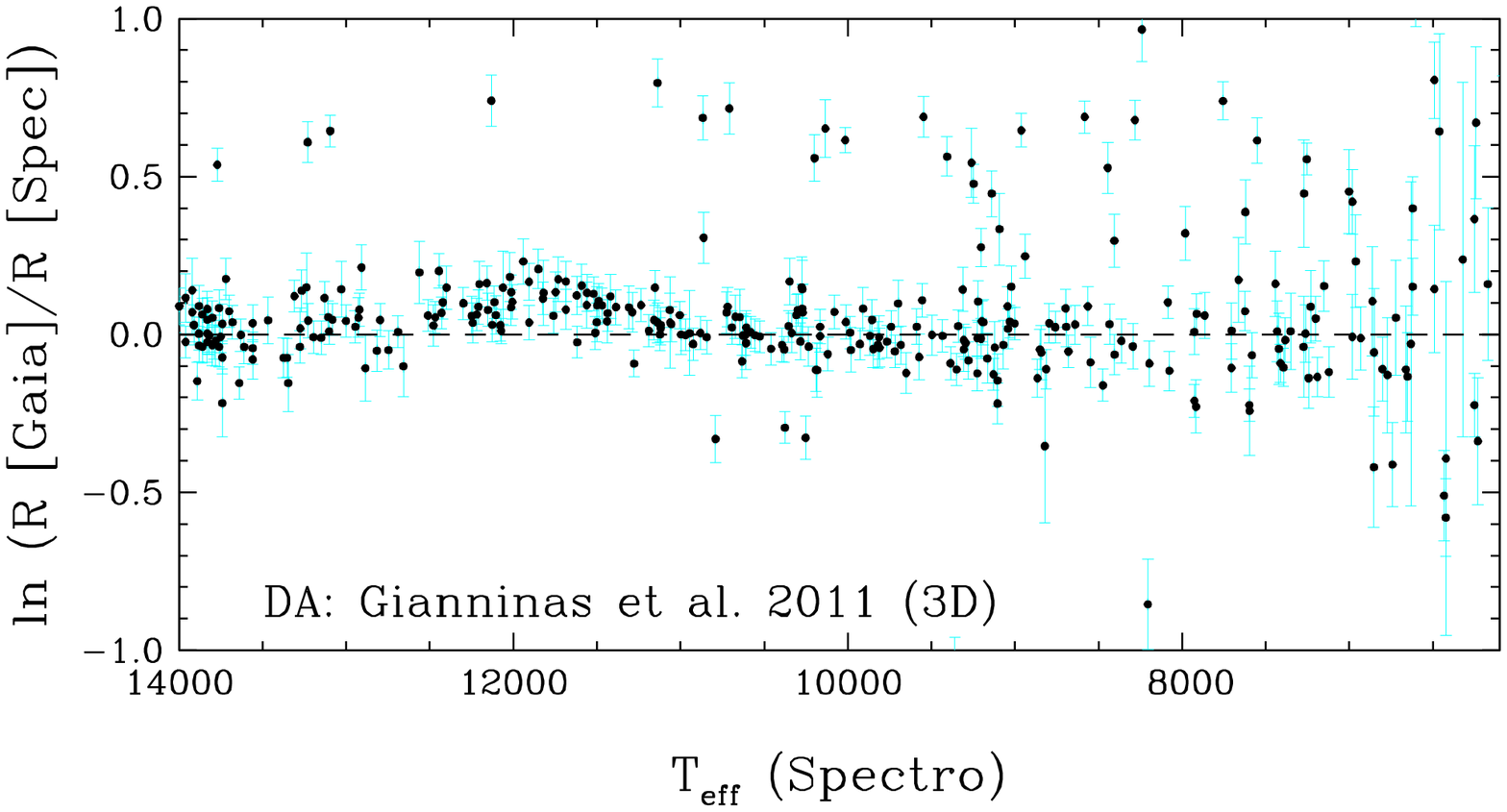}
\newline
\caption{Similar to Fig.~\ref{fig8} but for the radius ratio.\label{fig9}}
\end{figure}

Figs.~\ref{fig8}-\ref{fig9} highlight the ratio between photometric and spectroscopic parameters in the region where 3D effects are important for the \citet{gianninas11} sample. We employ units of $T_{\rm eff}^2$ and $R$, the latter directly derived from the surface gravity because we assume a mass-radius relation, as they are directly proportional to the predicted parallax (Section~\ref{sec3}). It is seen that 3D corrections mostly impact the predicted radii, and that there remains a small residual bump around 12\,000~K when comparing with \textit{Gaia}, in agreement with the inflexion observed at the same temperature in Section~\ref{sec3} for the predicted parallaxes. It is interesting to note that \citet{tremblay13} had already noticed a small residual bump, with an amplitude of $\approx$ 6\%, in the 3D mass distribution of the \citet{gianninas11} sample in the range between 11\,000~K and 12\,000~K. This is consistent with a slight residual high-mass problem that is not fully accounted by 3D $\log g$ corrections in that range. Finally, Fig.~\ref{fig9} confirms a tight sequence of white dwarfs where \textit{Gaia} suggests that the radius is larger by a factor of about two, or in other words that there are actually two DA white dwarfs at the observed location with similar $T_{\rm eff}$ and $\log g$.

\begin{figure*}
\centering
\includegraphics[width=1.0\columnwidth,bb = 15 195 590 590]{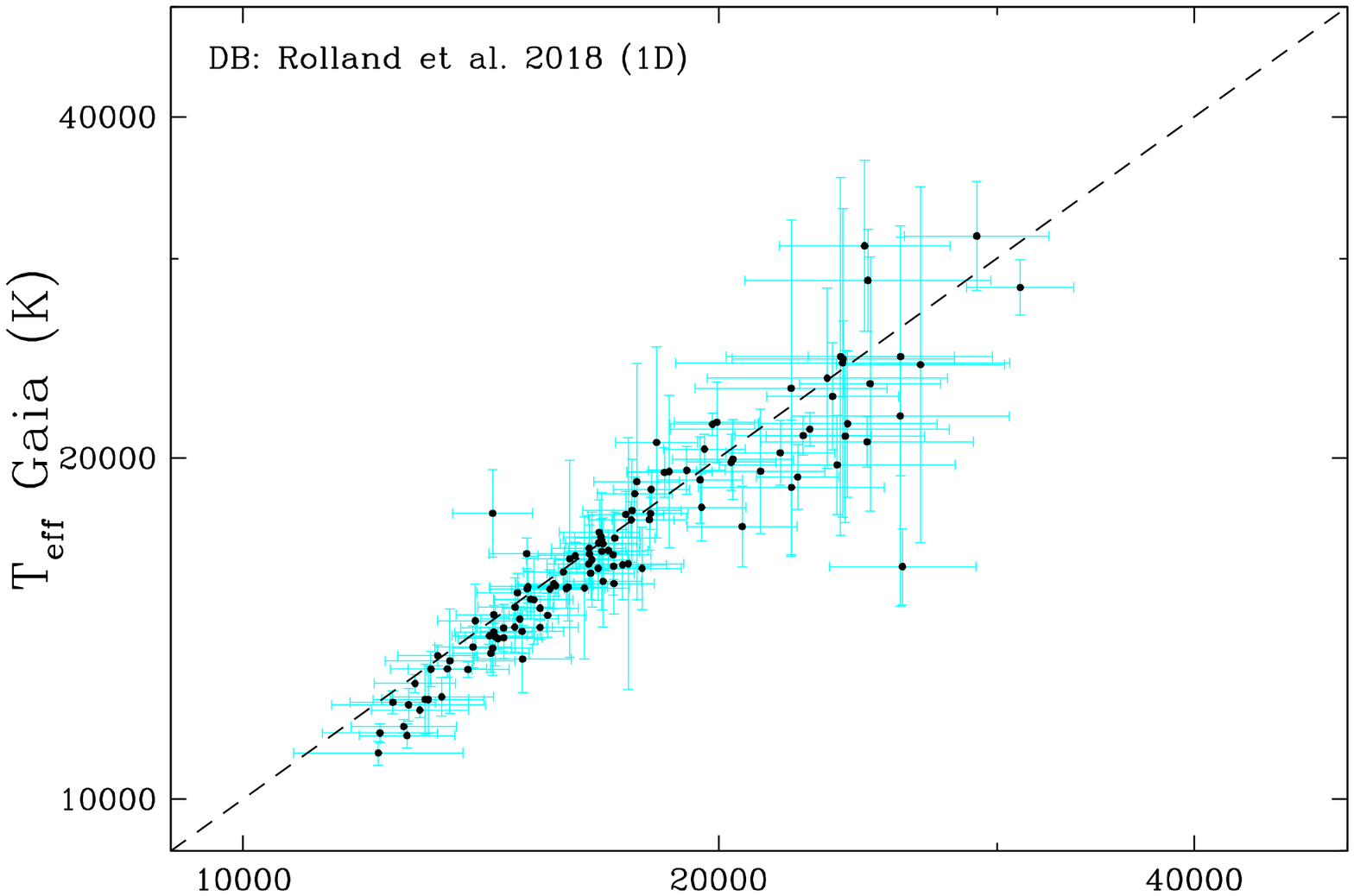}
\includegraphics[width=1.0\columnwidth,bb = 15 195 590 590]{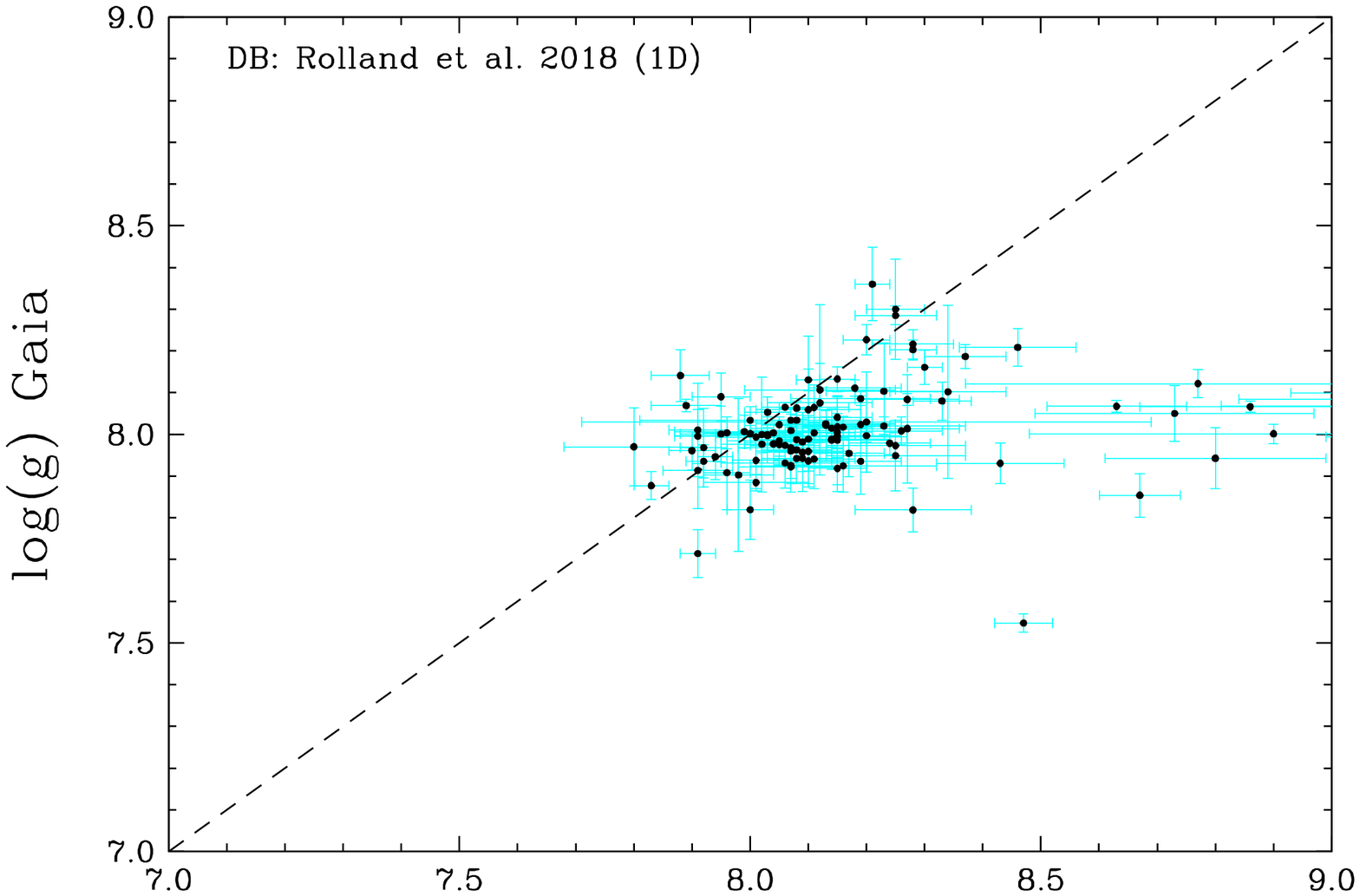}
\newline
\includegraphics[width=1.0\columnwidth,bb = 15 170 590 590]{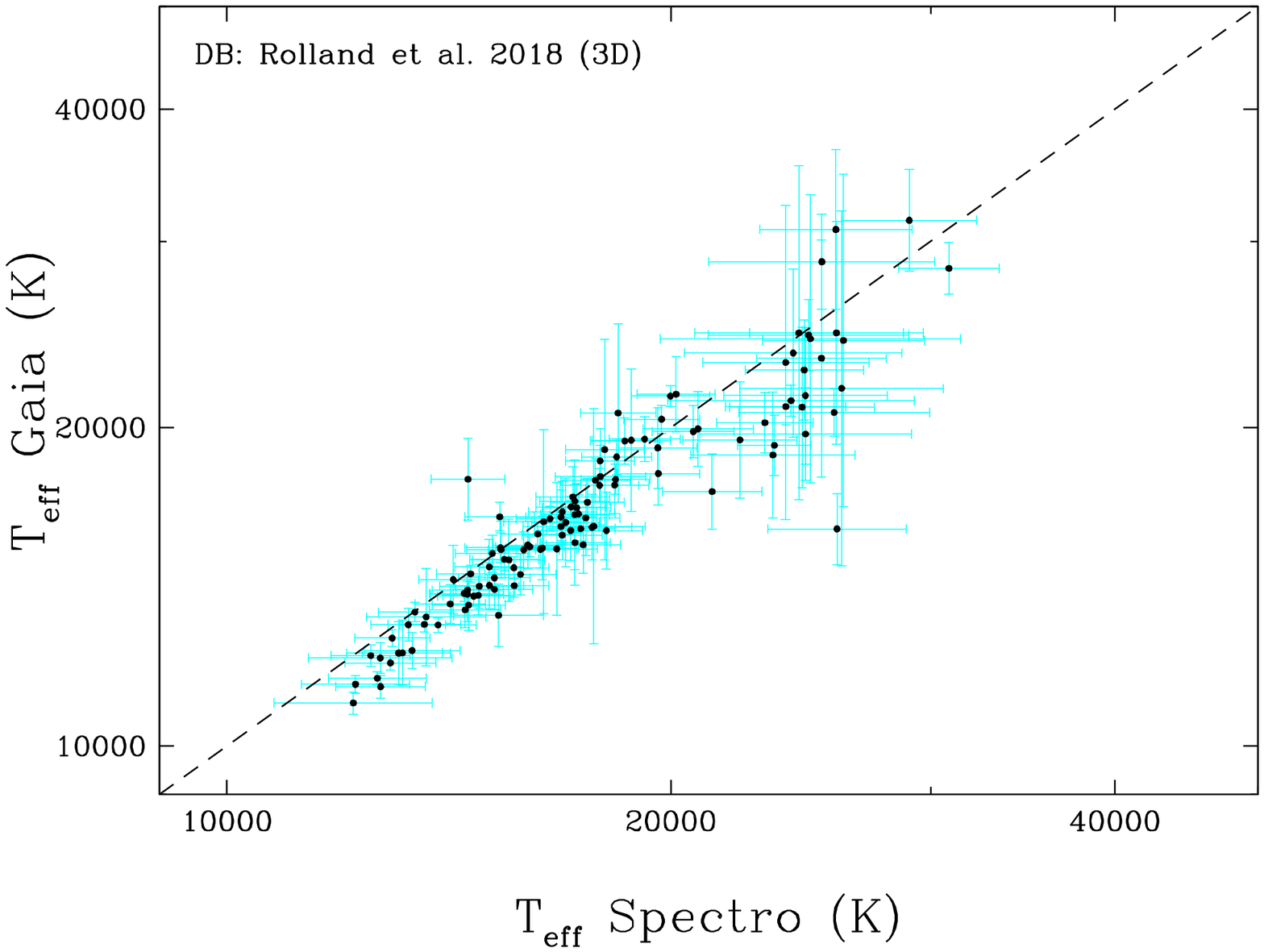}
\includegraphics[width=1.0\columnwidth,bb = 15 170 590 590]{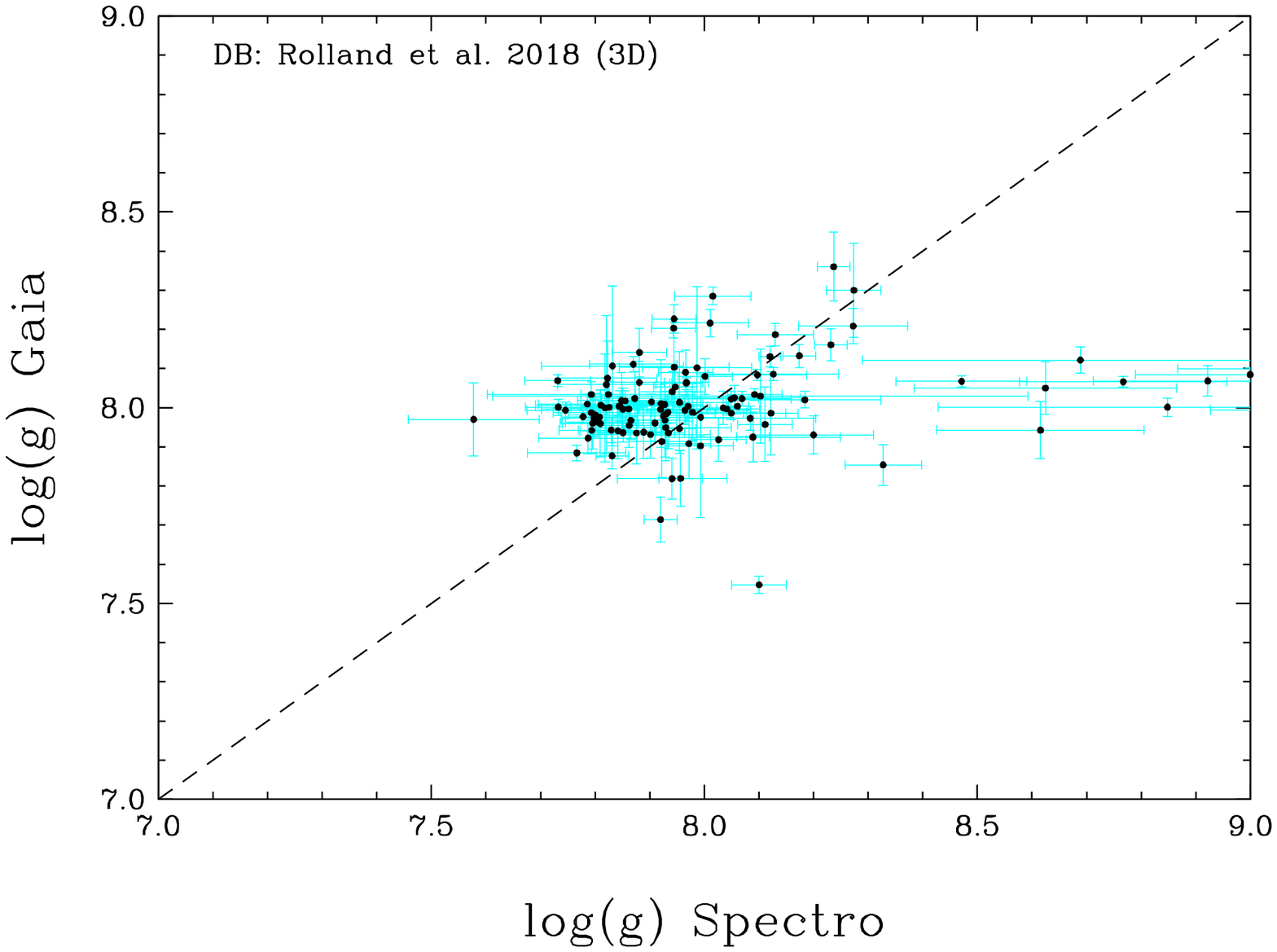}
\newline
\caption{{\it Top panels:} Comparison of spectroscopic and photometric \textit{Gaia} $T_{\rm eff}$ values (left panel) and surface gravities (right panels) for the DB and DBA sample of \citet{rolland18}. The one-to-one agreement between the two measurements is indicated by the dashed lines. {\it Bottom panels:} Similar to the top panels but with the 3D spectroscopic parameter corrections of \citet{cukanovaite18} assuming pure-helium atmospheres. \label{fig10}}
\end{figure*}

\begin{figure*}
\centering
\includegraphics[width=1.0\columnwidth,bb = 15 195 590 590]{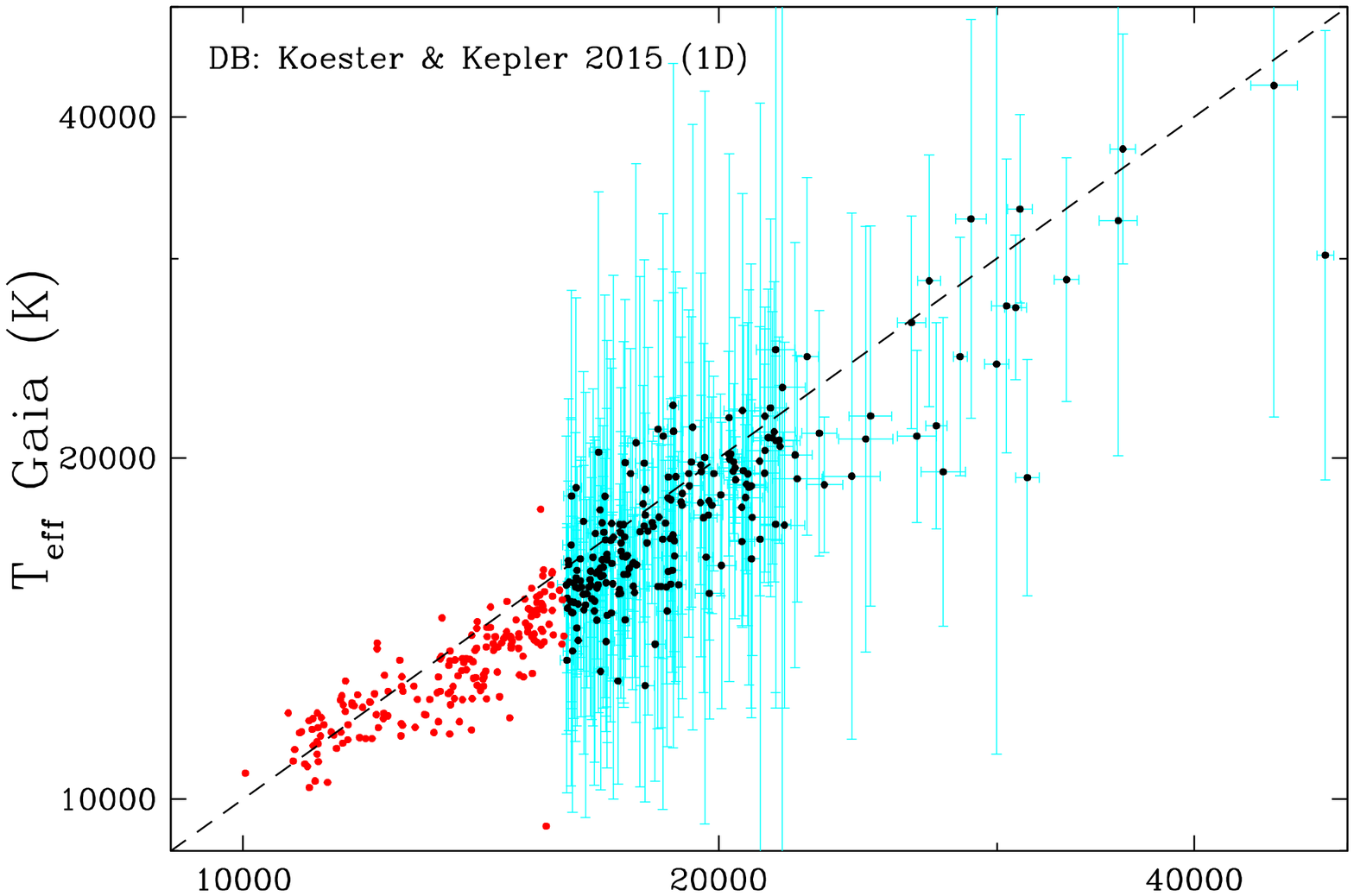}
\includegraphics[width=1.0\columnwidth,bb = 15 195 590 590]{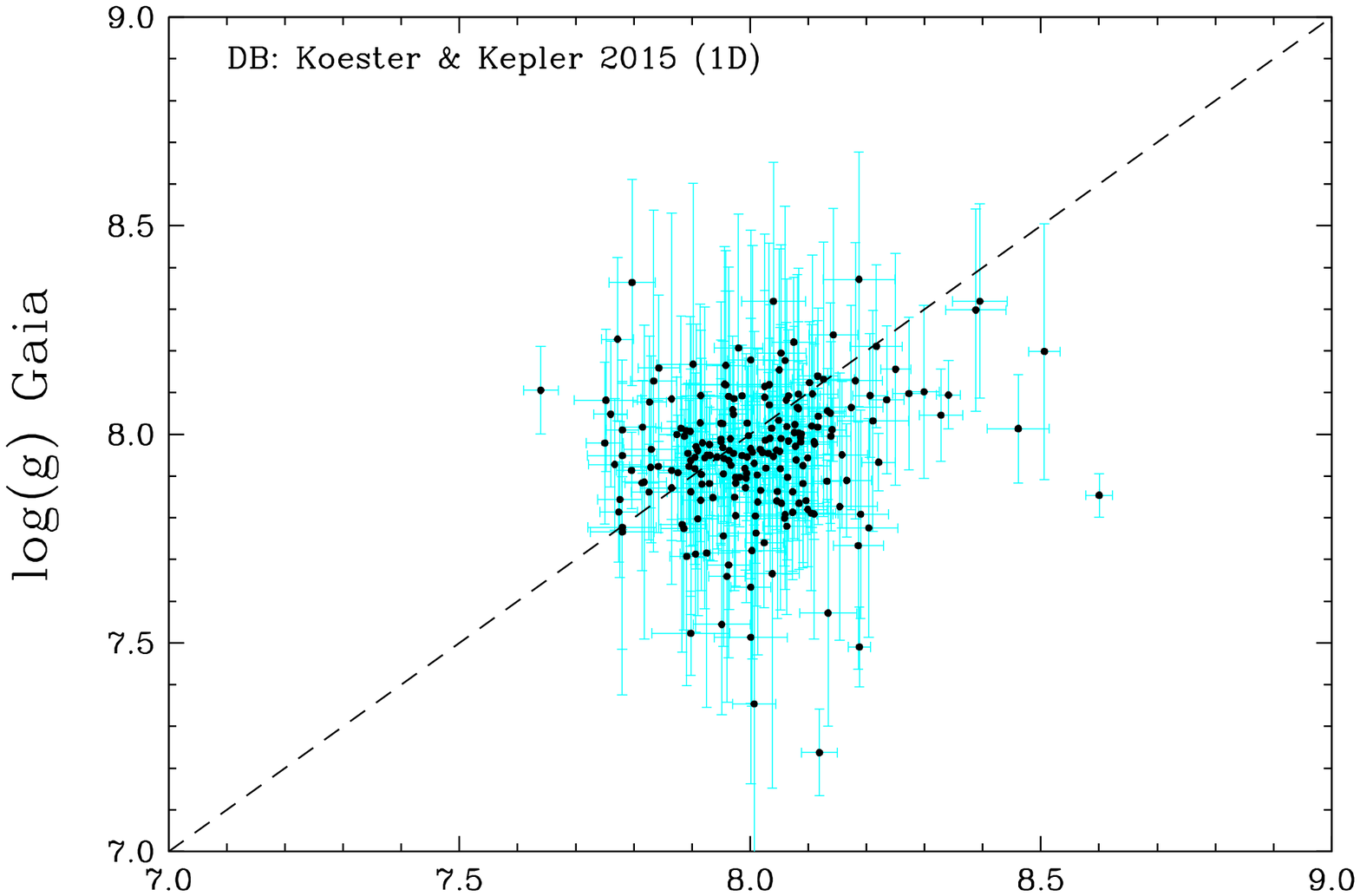}
\newline
\includegraphics[width=1.0\columnwidth,bb = 15 170 590 590]{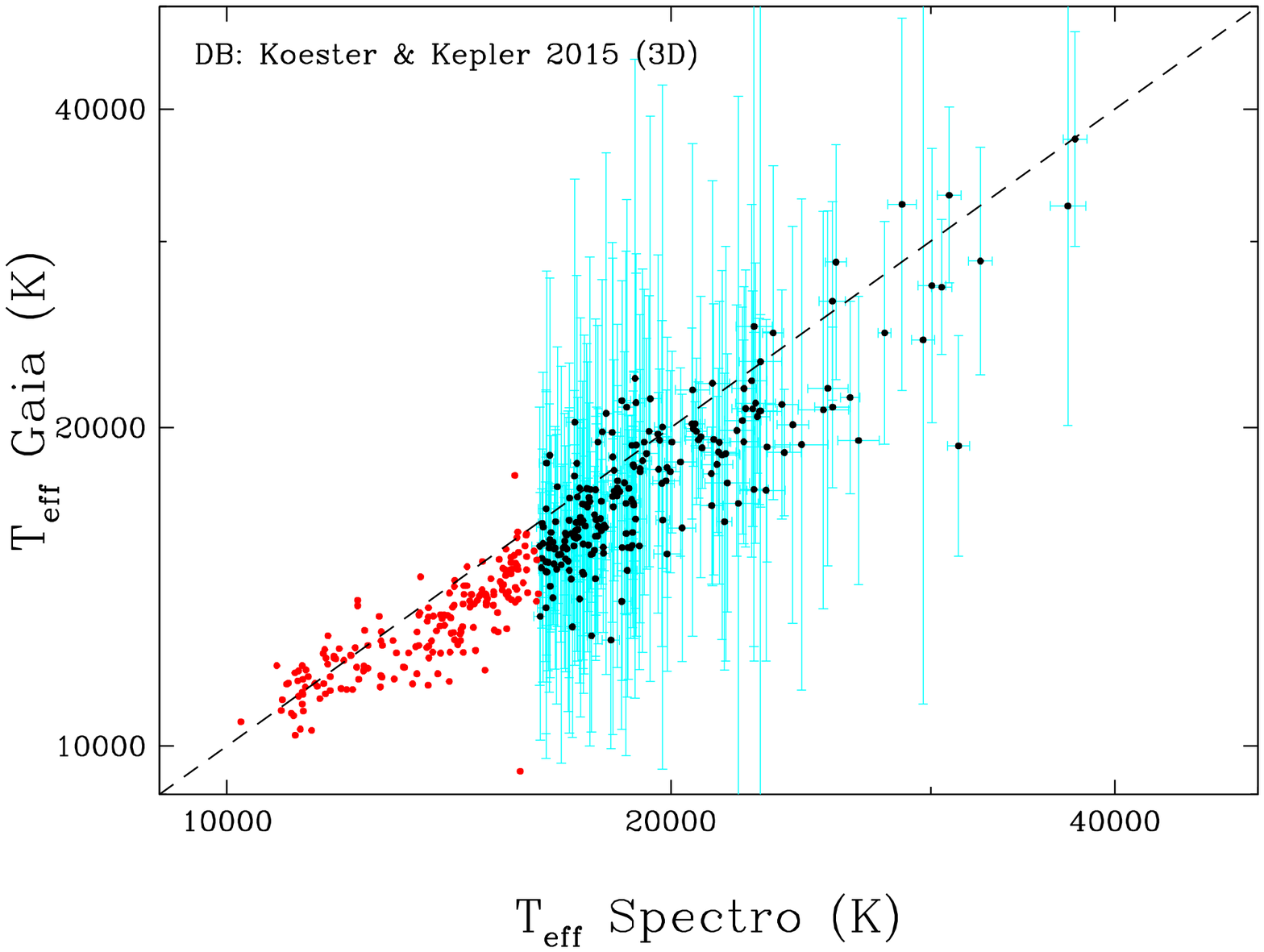}
\includegraphics[width=1.0\columnwidth,bb = 15 170 590 590]{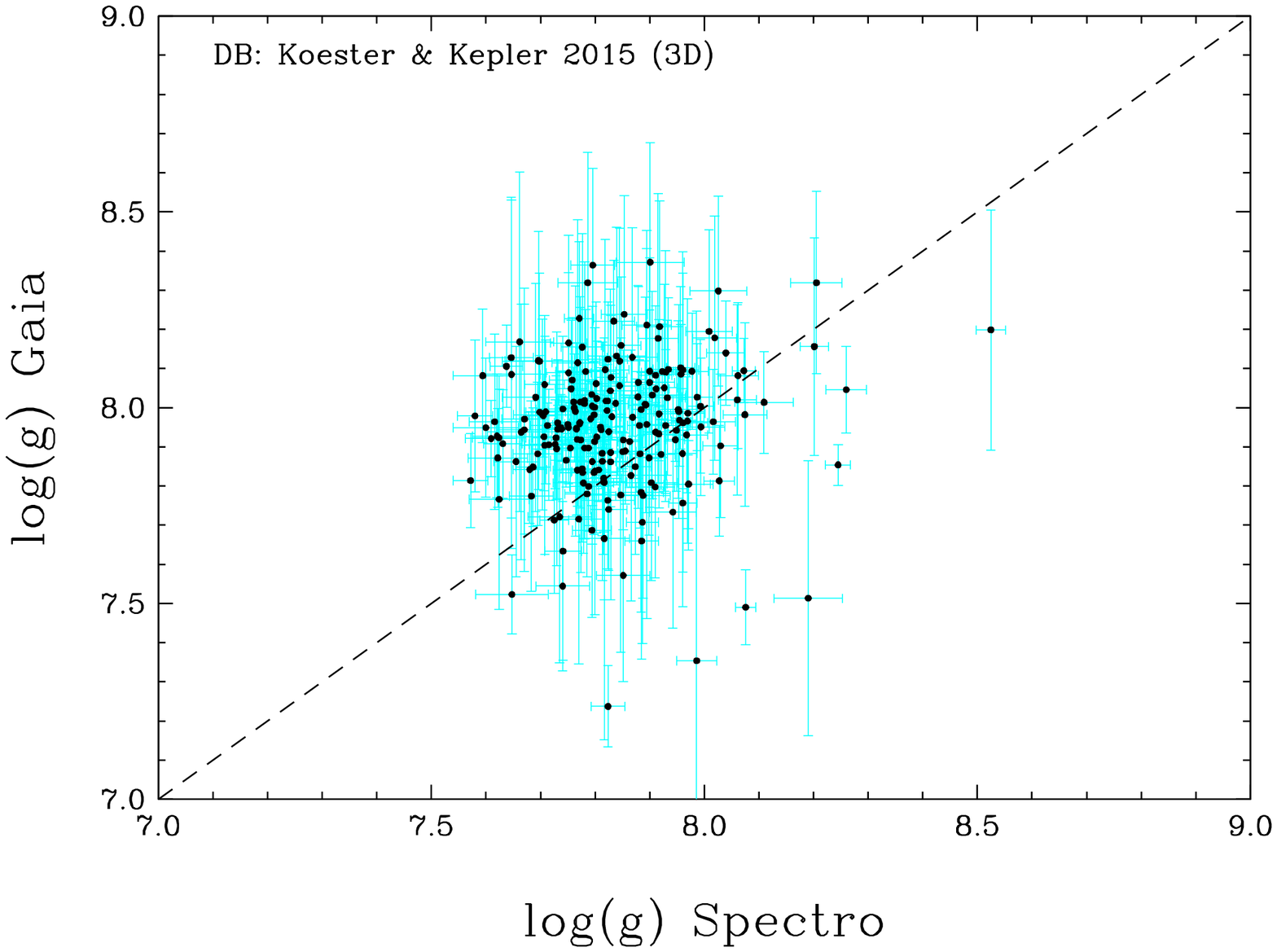}
\newline
\caption{Similar to Fig.~\ref{fig10} but for the DB/DBA sample of \citet{koester15}. Objects with fixed spectroscopic parameters at a value of $\log g = 8.0$ are plotted as red points with no error bars on the left panels and omitted from the comparison of $\log g$ values. \label{fig11}}
\end{figure*}

The comparisons of \textit{Gaia} photometric parameters with the DB/DBA spectroscopic samples of \citet{rolland18} and \citet{koester15} are presented in Figs.~\ref{fig10}-\ref{fig11}. We observe a pattern similar to that seen for DA white dwarfs, with \textit{Gaia} $T_{\rm eff}$ values being systematically smaller than the spectroscopic ones, both for 1D and 3D models. There is no evidence, however, that the cause is the same as that for DA stars. As it was observed in Section~\ref{sec3}, the 3D corrections of \citet{cukanovaite18}, assuming pure-helium atmospheres, have a significant effect and must be included in the comparison, although they do not bring the spectroscopic parameters into obviously better agreement with \textit{Gaia}. This must be re-assessed once 3D corrections for DBA stars become available. Finally, we note that all large spectroscopic $\log g$ values in the \citet{rolland18} sample, which correspond to objects with $T_{\rm eff} < 14\,000$~K, are disproved by \textit{Gaia}, suggesting instead standard $\log g = 8.0$ values for these stars. The discrepancy is not observed in the \citet{koester15} sample, as they artificially fixed the surface gravity to $\log g = 8.0$ for objects with $T_{\rm eff} < 16\,000$~K and these stars are not shown on the right panel of Fig.~\ref{fig11}.

\begin{figure*}
\centering
\includegraphics[width=1.0\columnwidth,bb = 15 160 500 590]{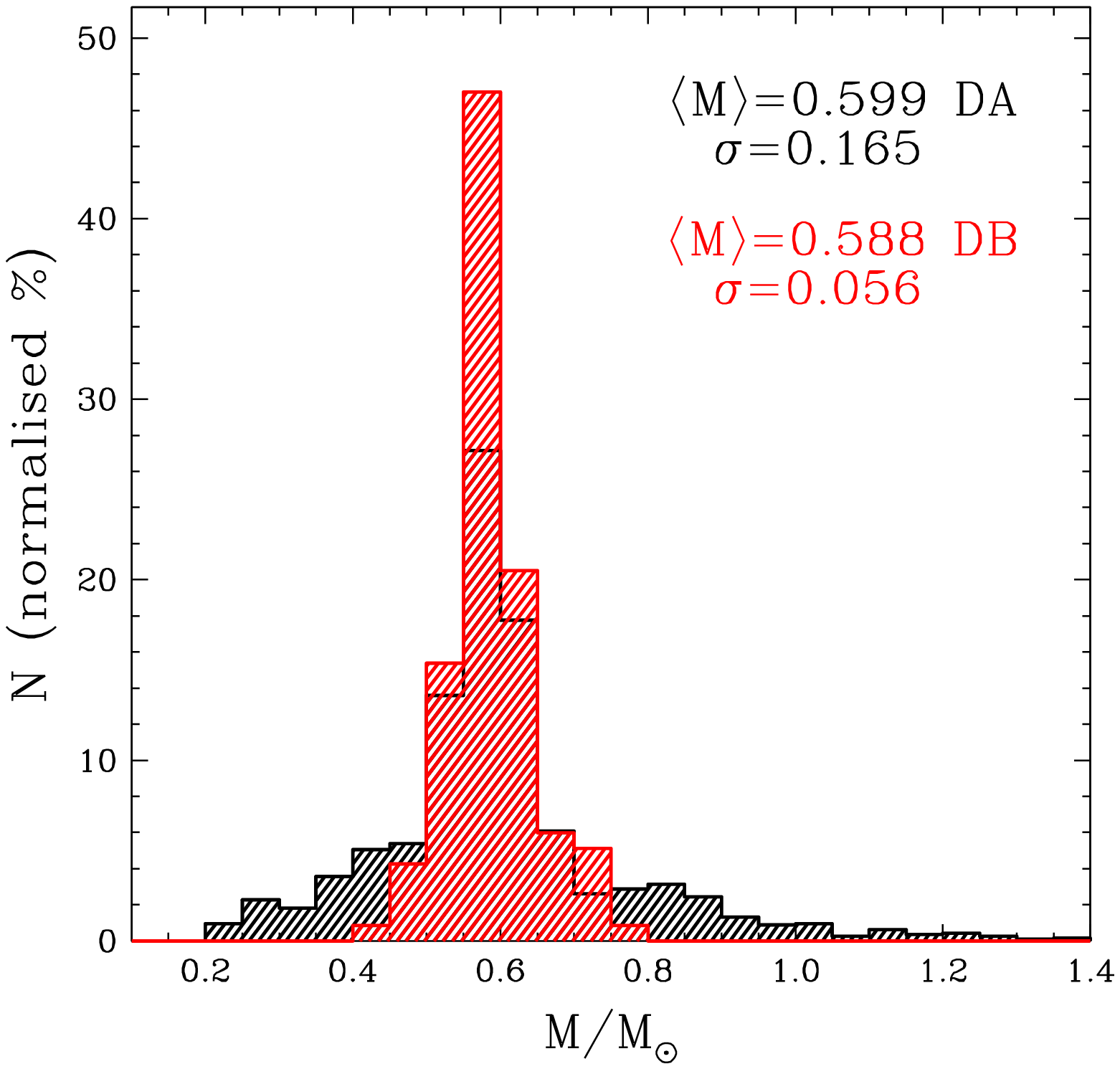}
\includegraphics[width=1.0\columnwidth,bb = 15 160 500 590]{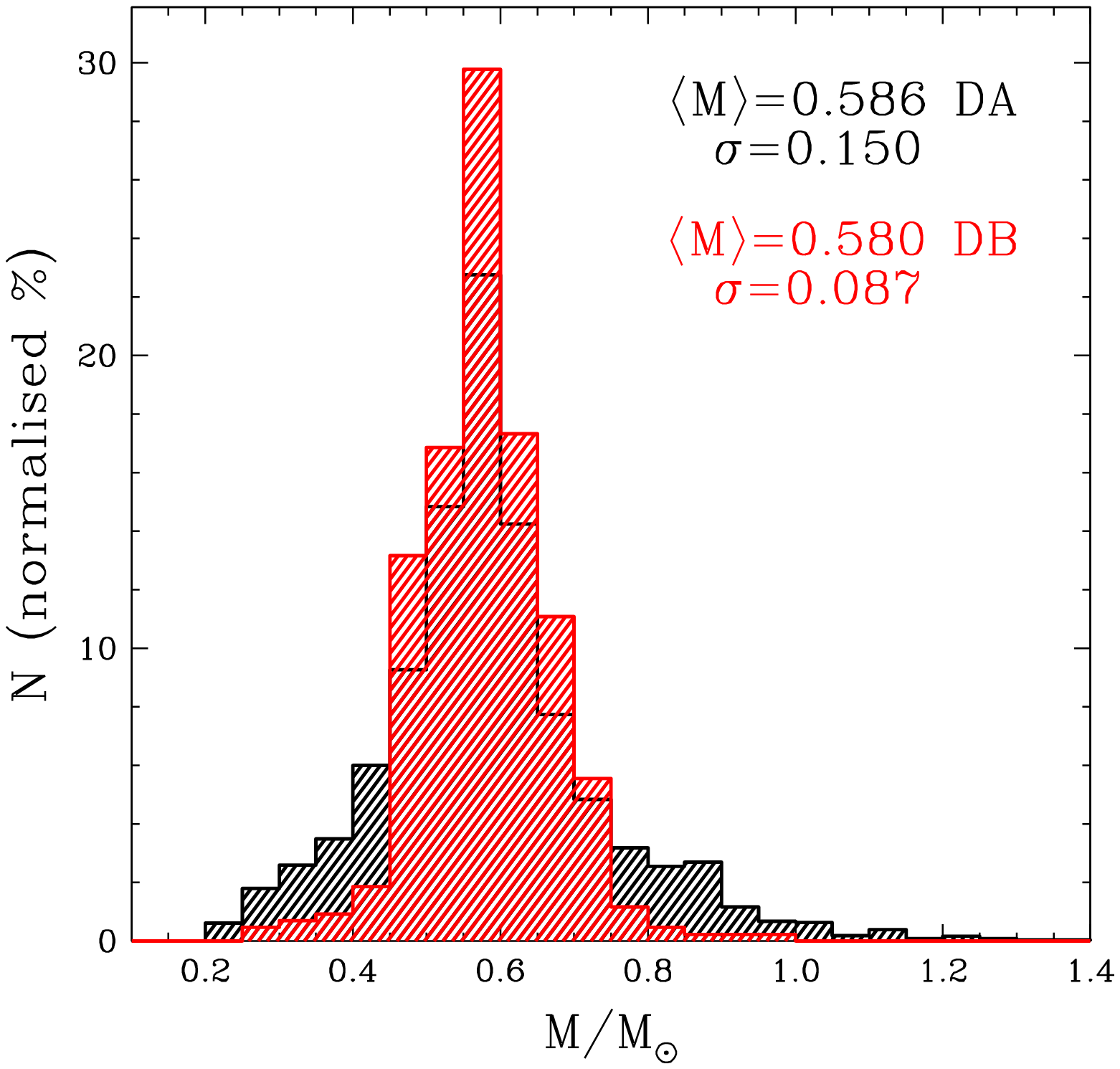}
\newline
\caption{{\it Left:} \textit{Gaia} photometric mass distributions for the DA white dwarfs in \citet{gianninas11} and the DB and DBA stars in \citet{rolland18}. The total numbers of white dwarfs have been renormalised to unity for an easier comparison of the different samples. {\it Right:} Similar to the left panel but for the DA white dwarfs in  SDSS DR7 and the DB/DBA stars from SDSS DR12 \citep{koester15}.\label{fig12}}
\end{figure*}

\section{Discussion}
\label{sec5}

\textit{Gaia} photometric parameters allow the derivation of stellar masses by adopting the white dwarf mass-radius relations described in Section~\ref{sec2}. Despite their incompleteness, our adopted samples are expected to present a reasonable picture of the field white dwarf mass distribution for a magnitude limited survey, similarly to what has been done in the pre-\textit{Gaia} era \citep[see, e.g.,][]{tremblay16}. \textit{Gaia} selected samples will be more appropriate in the future to study the astrophysical implications of the white dwarf mass distribution, but such samples currently have very poor spectroscopic completeness beyond 20\,pc, leading to potentially erroneous interpretations given the strong colour differences between cool DA and DC white dwarfs \citep{el-badry18,gentilefusillo18}.

Fig.~\ref{fig12} compares the photometric mass distributions of DA and DB white dwarfs from the samples of \citet{gianninas11} and \citet{rolland18}, and those found in the SDSS. We neglect the DA sample from SDSS-III BOSS as the results are very similar to SDSS DR7 and for DB stars we have made no such distinction between the SDSS spectrographs. We remind the reader that the mass distributions in Fig.~\ref{fig12} are independent from spectroscopy apart from the sample selection. We find very similar mean masses for all studied samples, suggesting that there is no strong bias from the selection function. 

The SDSS sample of DB white dwarfs clearly has a larger standard deviation around the mean compared to the more local \citet{rolland18} sample. It illustrates that observational scatter dominates over intrinsic width for the SDSS sample. This is a remarkable result, illustrating that the standard deviation in DB masses is at least three times smaller than for DA white dwarfs and it is unclear if even {\it Gaia} is able to resolve the DB mass distribution. For DA white dwarfs, the standard deviations agree well between samples and also between photometric and spectroscopic analyses \citep{gianninas11,tremblay11}, confirming that the mass distribution is well resolved.

\textit{Gaia} DR2 robustly suggests that the mean mass of DA and DB white dwarfs is the same within the uncertainties in the selection function of the samples and the assumption of thick hydrogen layers for all DA stars. However, there is also clear evidence that the DA and DB mass distributions have different shapes, with a sharp decline of the number of DB white dwarfs both at masses lower and higher than the average. These properties observed by \textit{Gaia} have been suggested before from spectroscopic analyses \citep{bergeron01,koester15,rolland18}, but they are now much more firmly established, with possible systematics from 3D effects and line broadening theories largely removed. This illustrates that while both spectral types may have the same mean mass, it can not be concluded that they originate from the same range of progenitors.

Almost all spectroscopically confirmed non-magnetic white dwarfs (e.g., excluding hot DQs, \citealt{dufourNat}) in the age range where DB stars are found  appear to have hydrogen-atmospheres for masses below $\approx$ 0.5 $M_{\odot}$ and above 0.7 $M_{\odot}$. This trend appears to continue for older and cooler remnants according to the \textit{Gaia}-SDSS catalogue of \citet{gentilefusillo18}. The behaviour at low masses is not surprising, as these objects are either low-mass white dwarfs ($M_{\rm WD} \lesssim$ 0.45~$M_{\rm \odot}$) or double degenerates. The former must have been formed through close binary evolution to have evolved in less than a Hubble time, and the samples of eclipsing binaries from \citet{parsons17} or the extremely-low-mass (ELM) white dwarfs from \citet{gianninas14} exclusively contain hydrogen-atmosphere stellar remnants. Double degenerates with mixed spectral types were likely eliminated from our samples as the optical spectra can identify them, which implies that only rare instances of DB+DB double degenerates  may populate the very low-mass tail of the DB mass distribution. 

The situation at higher-than-average masses is discussed in \citet{gentilefusillo18}, with a strong suggestion from young stellar clusters and \textit{Gaia} that massive white dwarfs overwhelmingly have thick hydrogen layers \citep{kalirai05-02} and are not subject to convective dilution or convective mixing \citep{rolland18}. Furthermore, the detailed shape of the DA mass distribution above 0.6 $M_{\rm \odot}$ could be explained from features in the initial-to-final-mass relation such as the onset of the second dredge-up \citep{tremblay16,el-badry18}. There is no evidence to rule out binary evolution for some of these objects \citep{toonen17}, but the latter does not lead to an unique, recognisable feature on the higher-than-average side of the white dwarf mass distribution.

\section{Summary}
\label{sec6}

\textit{Gaia} DR2 has detected about eight times more white dwarfs than previously known \citep{gentilefusillo18}. This allows for a rich variety of astrophysical applications, from constraining the local stellar formation history to the initial-to-final mass relation. As a first step, it is essential to understand the accuracy and systematics when deriving the fundamental parameters of these stellar remnants. 

We have used DA and DB white dwarfs with well understood spectroscopic parameters and mass-radius relations, and compared these to \textit{Gaia}-derived absolute magnitudes and fundamental stellar parameters. We emphasise that for  all samples studied in this work \citep[][as well as the SDSS]{gianninas11,rolland18}, individual spectroscopic $T_{\rm eff}$ and $\log g$ values of both DA and DB white dwarfs are generally in good agreement with {\it Gaia} within 2$\sigma$ when the assumption of a single star is verified. For a few percent of the objects, {\it Gaia} has clearly uncovered previously unsuspected double degenerates, although we made no attempt to fully separate the two populations from the available data because of poorly constrained selection functions.

We confirm that even with the precision of \textit{Gaia} DR2, it is not possible to robustly test the mass-radius relation or derive the hydrogen layer thickness on a star-by-star basis. We observe no evidence of a bimodal distribution of thin and thick layers, which is expected to be most prominent at large temperatures where a warm and thick hydrogen layer leads to a significant increase in stellar radius \citep{tremblay17}. However, these hot white dwarfs are generally distant, resulting in a lower \textit{Gaia} precision and additional uncertainties from reddening. Furthermore, it is possible that additional scatter owing to the presence of undetected metals and NLTE effects \citep{gianninas10} may hinder the detection of thin H-layers.

The size and precision of the \textit{Gaia} sample has allowed the detection of small systematic offsets between the spectroscopic and photometric parameters. There is a small residual bump in the spectroscopic $\log g$ distribution of DA white dwarfs in the temperature range from 11\,000 to 13\,000~K,  which is likely related to an incomplete account of 3D effects. For DB white dwarfs in the regime below $T_{\rm eff} < 14\,000$~K, we find that the spectroscopic technique fails to get accurate surface gravities as it has long been suspected \citep{bergeron11}, and {\it Gaia} photometric parameters do not appear to have such issue. This is consistent with a problem with neutral line broadening. For both DA and DB white dwarfs, the photometric and spectroscopic temperature scales appear to have a slight systematic offset, with {\it Gaia} systematically predicting slightly cooler temperatures by a few percent. We have no obvious explanation for this offset, but residual issues with line broadening theories in spectral analyses remain a possibility. 

The present study suggests that \textit{Gaia} and current mass-radius relations can be employed to derive precise $T_{\rm eff}$, $\log g$, masses, and radii, which is the first step to link white dwarfs to their progenitors. {\it Gaia} has clearly confirmed that DA and DB white dwarfs have the same mean mass within less than two percent. However, it was also found that the DB mass distribution has a significantly smaller intrinsic width around its mean than for DA white dwarfs. The next steps will be to understand the precision of the {\it Gaia} photometric parameters for other types of white dwarfs as well as gain a better understanding of the cooling rates, e.g. from stellar remnants in wide binaries or clusters.

\section*{Acknowledgements}
The research leading to these results has received funding from the European Research Council under the European Union's Horizon 2020 research and innovation programme n. 677706 (WD3D) and the European Union's Seventh Framework Programme (FP/2007- 2013) / ERC Grant Agreements n. 320964 (WDTracer).

This work has made use of data from the European Space Agency (ESA) mission {\it Gaia} (\url{https://www.cosmos.esa.int/gaia}), processed by the {\it Gaia} Data Processing and Analysis Consortium (DPAC, \url{https://www.cosmos.esa.int/web/gaia/dpac/consortium}). Funding for the DPAC has been provided by national institutions, in particular the institutions participating in the {\it Gaia} Multilateral Agreement. 

\bibliographystyle{mnras}
\bibliography{aamnem99,aabib_tremblay}

\cleardoublepage
\onecolumn
\appendix
\section{\textit{Gaia}-SDSS spectroscopic fits}
\label{app_A}

\scriptsize
\begin{longtable}{lllllllllll}
\caption{\label{A1} Subsample of \textit{Gaia}-SDSS DR1-DR7 DA white dwarfs} \\
\hline
\hline
SDSSJ name &  MJD-plate-fiber & $\pi_{\rm Gaia}$ & $\sigma_{\rm \pi~Gaia}$ & 3D $T_{\rm eff}$ & $\sigma_{\rm Teff}$ & 3D $\log g$ & $\sigma_{\rm logg}$ & $\pi_{\rm spectro}$ & $\sigma_{\rm \pi ~spectro}$ & Comment \\
 &   & [mas] & [mas] & [K] & [K] & [cgs] & [cgs] & [mas] & [mas] & \\
\hline
000022.88$-$000635.7 & 51791-0387-166 &   2.41460  &  0.28810  &   23013  &     473   &   7.443   &   0.064  &  1.84623 &   0.12224 \\
000034.06$-$052922.4 & 54380-2624-261 &   5.12254  &  0.17227  &   20299  &     148   &   7.799    &  0.022  &  5.23086 &   0.10821 \\
000034.07$-$010820.0 & 52203-0685-187 &   5.68123  &  0.23765  &   13006  &     222   &   8.025     & 0.054  &  5.59269 &   0.25335 \\
000051.85+272405.3 & 54452-2824-272 &   2.17536  &  0.38865  &   21961  &     371  &   7.946     & 0.051  &  2.25211 &   0.10966 \\
000100.42$-$042742.9 & 54327-2630-359 &   2.41227  &  0.32521  &   17138  &     225   &   7.604   &   0.041  &  2.50832 &   0.09545 \\
\hline
    \caption*{Notes: Atmospheric parameters are from Balmer line fitting and objects fitted with models including CNO \citep{gianninas10} are indicated in the Comment column. For objects with multiple spectra, we only kept the spectrum with the highest S/N. This is a portion of the table and the full data is available as Supplementary material (online).
    }
\end{longtable}

\begin{longtable}{lllllllllll}
\caption{\label{A2} Subsample of \textit{Gaia}-SDSS-III (BOSS) DR9-DR14 DA white dwarfs} \\
\hline
\hline
SDSSJ name &  MJD-plate-fiber & $\pi_{\rm Gaia}$ & $\sigma_{\rm \pi~Gaia}$ & 3D $T_{\rm eff}$ & $\sigma_{\rm Teff}$ & 3D $\log g$ & $\sigma_{\rm logg}$ & $\pi_{\rm spectro}$ & $\sigma_{\rm \pi ~spectro}$ & Comment \\
 &   & [mas] & [mas] & [K] & [K] & [cgs] & [cgs] & [mas] & [mas] & \\
\hline
000006.75$-$004653.9 & 56956-7850-0719   & 4.67747   & 0.35240     &11309    &  139     & 8.071    &  0.052  &  4.13786  &  0.19743\\
000022.87$-$000635.6 & 56959-7848-0062   & 2.41460   & 0.28810     &22497     &  185     & 7.492    &  0.025  &  1.90322  &  0.04927\\
000034.09$-$052922.4 & 56564-7034-0336   & 5.12254   & 0.17227     &19983     &  131     & 7.785    &  0.020  &  5.15266  &  0.09676\\
000104.05+000355.8 & 56959-7848-0026   & 3.53414   & 0.34839     &13126     &  157     & 8.100    &  0.036  &  3.54205  &  0.11033\\
000106.93+082825.5 & 55863-4534-0466   & 6.60360   & 0.30072      &7682      &  62      & 7.931    &  0.097  &  6.88215  &  0.48615\\

\hline
    \caption*{Notes: Atmospheric parameters are from Balmer line fitting and objects fitted with models including CNO \citep{gianninas10} are indicated in the Comment column. For objects with multiple spectra, we only kept the spectrum with the highest S/N. This is a portion of the table and the full data is available as Supplementary material (online).}
\end{longtable}

\end{document}